\documentclass[letterpaper,11pt]{article}
\usepackage[left=2.5cm,right=2.5cm,top=2.75cm,bottom=2.5cm,nohead]{geometry}
\pdfoutput=1

\usepackage{color,graphicx}

\usepackage[breakable]{tcolorbox}

\def\footnote#1{{}}

\def\noop#1{{}}
\def\yesir#1{{}}
\def\maybe#1{{}}
\def\nosir#1{{}}

\def\colorred#1{{\color{red}#1}}
\def\colorblu#1{{\color{blue}#1}}

\def\urlh#1{{}}

\def\emdash{---}
\def\etal{{\em{et al}}}

\makeatletter
\def\footnotextnonum{\xdef\@thefnmark{}\@footnotetext}
\makeatother

\title{Biological Blueprints for Next Generation AI Systems}

\date{}

\author{Thomas Dean$^{1,2}$\\
Chaofei Fan$^{2}$\\
Francis E. Lewis$^{2}$\\
Megumi Sano$^{2}$}\footnotextnonum{Affiliations: $^{1}$Brown University, $^{2}$Stanford University}

\begin{document}





\pagenumbering{gobble}

\begin{titlepage}

  \maketitle

  \begin{abstract}
    Diverse subfields of neuroscience have enriched artificial intelligence for many decades. With recent advances in machine learning and artificial neural networks, many neuroscientists are partnering with AI researchers and machine learning experts to analyze data and construct models. This paper attempts to demonstrate the value of such collaborations by providing examples of how insights derived from neuroscience research are helping to develop new machine learning algorithms and artificial neural network architectures. We survey the relevant neuroscience necessary to appreciate these insights and then describe how we can translate our current understanding of the relevant neurobiology into algorithmic techniques and architectural designs. Finally, we characterize some of the major challenges facing current AI technology and suggest avenues for overcoming these challenges that draw upon research in developmental and comparative cognitive neuroscience.
  \end{abstract}

\end{titlepage}

\newpage

\pagenumbering{roman}

\tableofcontents

\newpage

\pagenumbering{arabic}




\section{Introduction}



Artificial neural networks support distributed computations in which concepts are represented as patterns of activity in the units that comprise the network layers, and inference is carried out by propagating activation levels between layers weighted by learned connection weights.  Artificial neural networks provide a type of fast, flexible computing well suited to handling ambiguity of the sort we routinely encounter in real-world environments, and, by doing so, they complement traditional symbolic computing technologies.

Engineers frequently borrow ideas from nature and generally find it more practical to translate these ideas into current technology rather than attempt to reproduce nature's solutions in detail. Indeed, the basic idea of artificial neural networks has been implemented multiple times using different technologies in order to approximate the connectivity patterns and signal transmission characteristics of real neural circuits while largely ignoring the physiology of real neurons in their implementation.

The human brain supports a wide array of learning and memory systems. Some we have begun to understand functionally and behaviorally, others we can only hypothesize must exist, and still others about which we haven't a clue. Just knowing {\it{that}} the brain supports a particular capability can serve as an important clue in engineering complex AI systems. Knowing {\it{how}} can lead to an innovative design, enhanced performance and extended competence. In particular, knowing something about how specific biological circuits relate to behavior helps in designing novel network architectures.

We are interested in designing neural network architectures that leverage what is known about biological information processing to solve complex real-world problems. To focus our efforts, we have set out to design end-to-end systems that assist human programmers in writing, debugging and modifying software. We benefit considerably from working closely with scientists from diverse subdisciplines of neuroscience to seek solutions to specific problems and identify additional problems we may have overlooked. The following section explains why this commingling of people, ideas and technologies is so valuable to us in pursuit of our objectives.

\section{Neuroscience}
\label{section_neuroscience}


  

From the brain of an Etruscan shrew weighing in at less than a tenth of a gram to a sperm whale brain weighing more than eight kilograms, it is clear that natural selection has stumbled on a basic brain plan and set of developmental strategies that enables it to construct a diverse set of special-purpose brain architectures for efficiently expressing a wide range of sophisticated behavior~\cite{DouglasandMartinCURRENT-BIOLOGY-12,WillemetBRAIN-SCIENCE-12}. The human brain with its approximately 100 billion neurons and the shrew brain with approximately 1 million neurons share the same basic architecture.

The mouse brain has homologues of most human subcortical nuclei and has contributed significantly to our understanding of the human brain and human neurodegenerative disease in particular. The differences between between human and chimpanzee brains are subtle~\cite{Mora-BermudezetalELIFE-16} and yet humans display a much wider range of behavior and express a much larger repertoire of genes than any other species~\cite{HawrylyczetalNATURE-NEUROSCIENCE-15}. So what makes the difference?

It's the connections between neurons that matter or, more generally, it's the different types of communication between neurons that biologists refer to as {\it{pathways}}. There are electrical, chemical and genetic pathways and each of them obey different constraints and are used for different purposes. They include point-to-point and broadcast methods of communication~\cite{HanetalNATURE-18}. They transfer information at different speeds and using different coding strategies. Layered architectures are common not just in the cortex but throughout the brain. It's the wiring that sets humans apart.


\subsection{Connectivity}


\begin{figure}
  \begin{center} 
    \includegraphics[height=200pt]{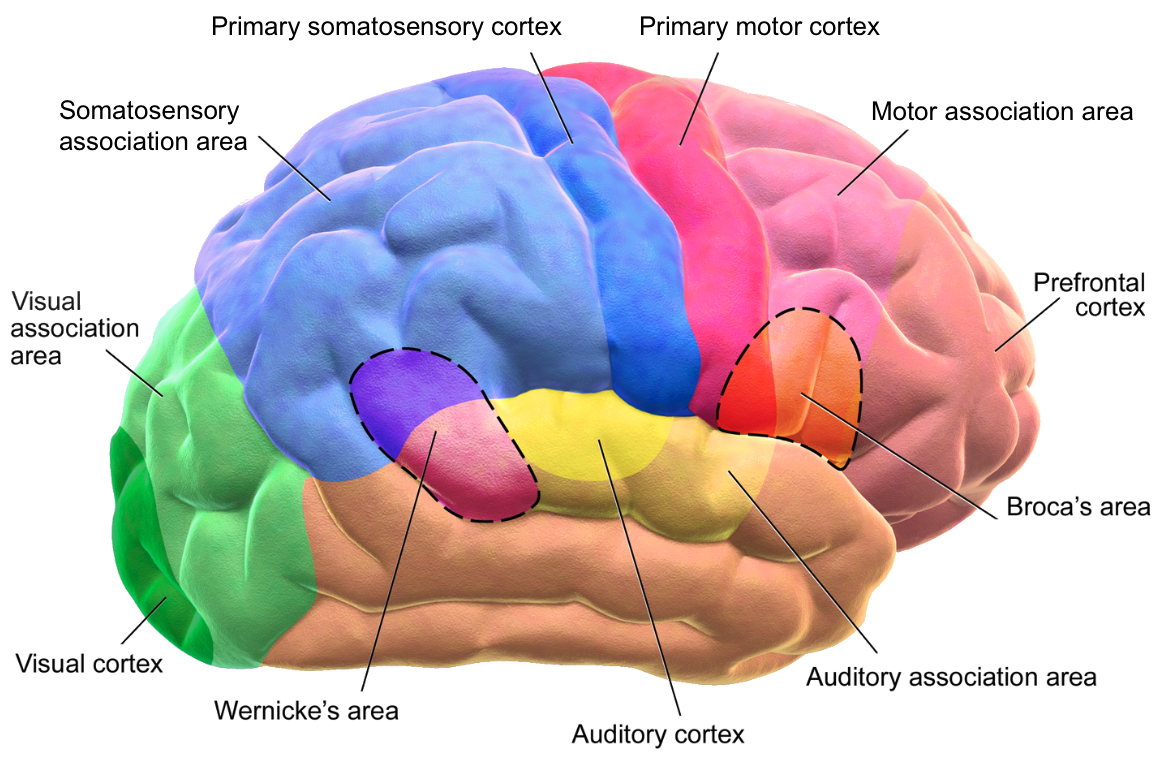} 
  \end{center}
  \caption{A highly stylized rendering of the major functional areas of the human cortex shown from the side with the head facing to right. Highlighted regions include the occipital lobe shown in shades of green including the primary visual cortex; the parietal lobe shown in shades of blue, including the primary somatosensory cortex; the temporal lobe shown in shades of yellow including the primary auditory cortex; and the frontal lobe shown in shades of pink, including the primary motor and prefrontal cortex. The region outlined by a dashed line on the right is Broca’s area and it is historically associated with the production of speech and hence its position relative to the motor cortex. The region outlined by a dashed line on the left is Wernicke’s area and it is historically associated with the understanding of speech and hence its position relative to the sensory cortex.}
  \label{fig_necortex}
\end{figure}



Figure~{\urlh{#fig_Human_Brain_Neocortex_Function}{\ref{fig_necortex}}} shows the major functional areas of the human neocortex including the primary and secondary sensory and motor areas. The proximal locations of the areas provide a very rough idea of how different functions might relate to another. The graphic shown belies the fact that the brain is three dimensional and much of its functional circuitry hidden under the cortical sheet. The human cerebral cortex is complexly folded to fit within the skull with much of it hidden within the folds. This folded sheet of tissue accounts for more than 75\% of the human brain by volume~\cite{SwansonTiN-95} and is largely responsible for the rich behavioral repertoire that humans exhibit. It is worth pointing out in this context that the cortical sheet enshrouds a collection of evolutionarily preserved and highly specialized circuits homologues of which are found in all mammals and without which the cortex would be useless.

The graphic shown in Figure~{\urlh{#fig_Human_Brain_Neocortex_Function}{\ref{fig_necortex}}} is a simplification of the standard medical textbook diagram. In particular, several of the association areas are not shown and those that are shown are labeled somewhat differently than is common practice. The organizing biological principle is that, the further away from the primary sensory areas, associative functions become more general by integrating information from multiple modalities to construct abstract representations tailored to serve ecologically relevant objectives~\cite{Higher_Cortical_Functions_Association}. It is worth contemplating the arrangement of areas to note that they converge on locations in the cortex that will play an important role in decision making and higher-order cognition more generally.


\begin{figure}
  \begin{center} 
    \includegraphics[height=150pt]{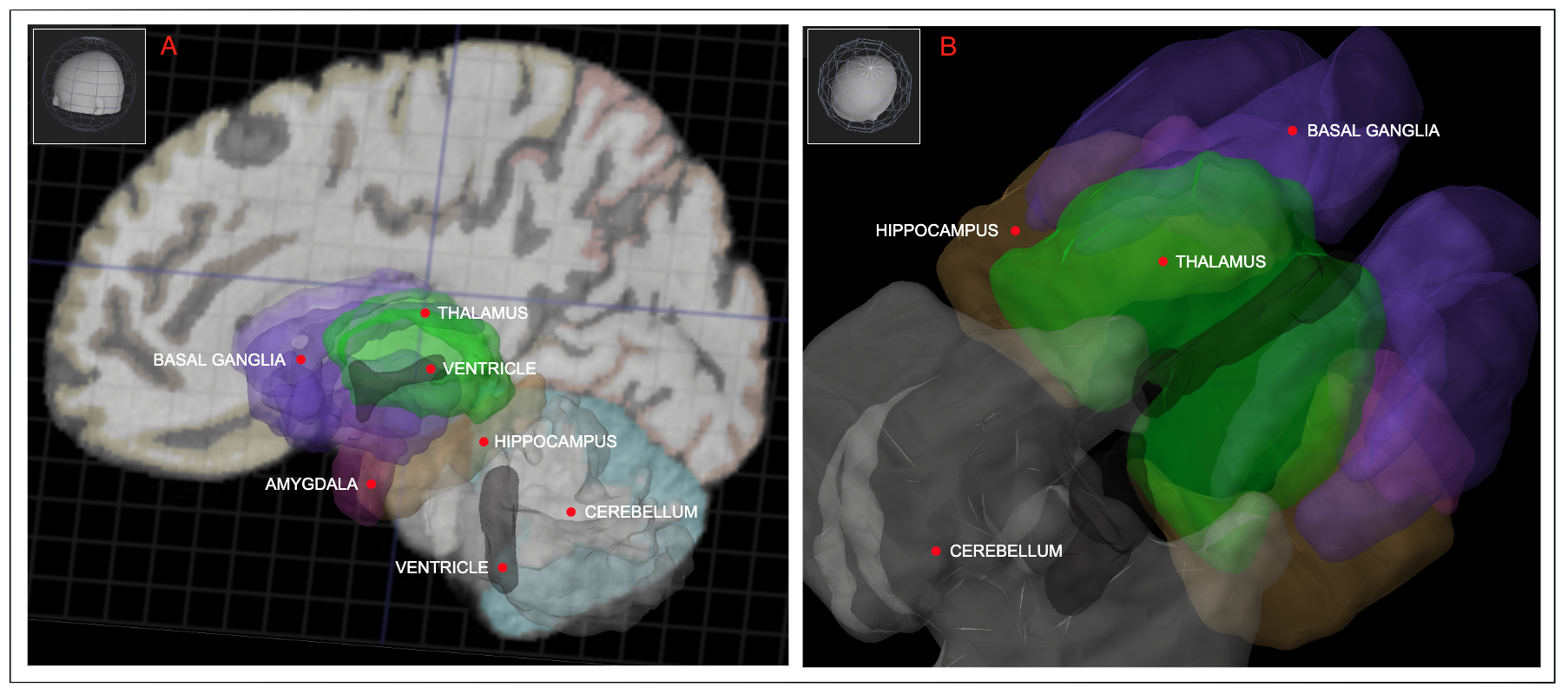} 
  \end{center}
  \caption{Two 3-D renderings of the human brain generated by the Allen Institute {\urlh{ttp://human.brain-map.org/static/brainexplorer}{Brain Explorer}} from the Allen Human Brain Reference Atlas~\cite{HawrylyczetalNATURE-12}. The inset shown in the left upper corner of each panel indicates the orientation of the head. The left panel ({\colorred{A}}) shows 3-D reconstructions of several subcortical nuclei featured in this paper. A cross-sectional view of the whole brain is projected on the mid-sagittal plane dividing the right and left sides of the brain illustrating how the cortex envelopes the subcortical regions. The right panel ({\colorred{B}}) shows the same subcortical nuclei as seen from above (horizontal plane) and to the rear of the brain illustrating how the thalamus is located between the cortical sheet and the subcortical nuclei serving in its role as a relay between the two regions.}
  \label{fig_brains}
\end{figure}


Figure~{\urlh{#fig_Human_Brain_Atlas_Allen_Institute}{\ref{fig_brains}}} highlights the 3-D structure of several subcortical nuclei emphasized in this paper showing how they relate anatomically to one another and to the cortex. The reconstructions were generated from data generated by {\it{functional magnetic resonance imaging}} (fMRI) of adult human subjects~\cite{HawrylyczetalNATURE-12} and offer additional functional insight complementing conventional histological studies~\cite{BridgeandClarePTRS-B-06}. They don't however provide detailed information concerning either local or long-range connectivity.

Traditionally, tracing connections between individual neurons has been accomplished using a variety of techniques including conventional histological and staining techniques, electrophysiology, neurotropic retroviruses and transgenic organisms expressing fluorescent proteins. However, these methods yield relatively sparse reconstructions and don't scale well to large tissue samples~\cite{Arenkiel2014neural,CallawayCURRENT-OPINION-08}.

Small samples of neural tissue can be fixed, stained and sliced into thin sections. Each of the sections is then scanned with an electron microscope and the resulting digital images analyzed with computer vision software to reconstruct neurons and identify synapses~\cite{MikulaandDenkNATURE-METHODS-15}. The process is time consuming but can be fully automated and scaled to handle larger samples~\cite{JanuszewskietalNATURE-METHODS-18,ZhengetalCELL-18}. 

It is also possible to reconstruct the major {\it{white matter tracts}} formed by bundles of myelinated fibers using diffusion-weighted fMRI and averaging over multiple subjects after registering the individual brain scans with a reference atlas~\cite{OishietalNEUROIMAGE-08,WakanaetalRADIOLOGY-04}. Unlike the previous technologies, this method is not destructive so it can be applied to human subjects and accuracy is improved by averaging over multiple subjects after registering the individual brain scans with a reference atlas

These major tracts increase the speed of signal transmission between regions allowing for more distant communication in larger brains. The differences between the neocortex in humans and chimpanzees are subtle~\cite{Mora-BermudezetalELIFE-16}; however, white matter connections observed in humans but not in chimpanzees particularly link multimodal areas of the temporal, lateral parietal, and inferior frontal cortices, including tracts important for language processing~\cite{ArdeschetalPNAS-19,Gomez-RoblesetalPNAS-15}.


\begin{figure}
  \begin{center} 
    \includegraphics[height=150pt]{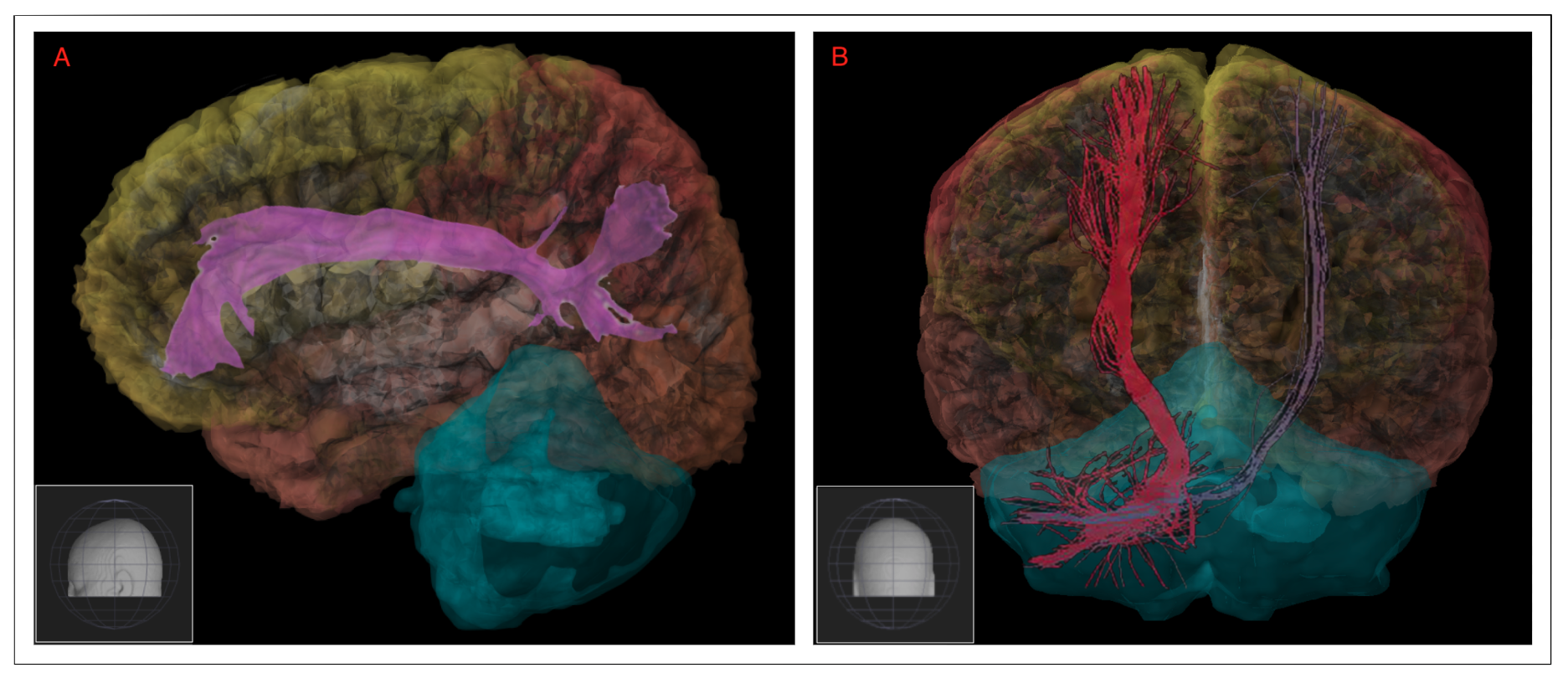} 
  \end{center}
  \caption{White matter tracts corresponding to bundles of myelinated neurons speed the transmission of information between distant regions of the brain. The left panel ({\colorred{A}}) shows the connections between the prefrontal cortex and circuits in the parietal and temporal cortex that shape conscious awareness, guide attention and support short-term memory maintenance~\cite{ChicaetalBSF-18,Dehaene2014}. The parietal and temporal cortices are known for being home to {\it{association areas}} that integrate information from multiple sensory systems thereby creating rich representations necessary for abstract thinking. In humans, white matter tracts between the frontal cortex and the cerebellum {\emdash{}} shown in the right panel ({\colorred{B}}) {\emdash{}} facilitate higher-order cognitive function in addition to their role in supporting motor function in all mammals. For example, these connections are particularly important in the development of reading skills in children and adolescents~\cite{TravisetalHBM-15,KozioletalCEREBELLUM-14}.}
  \label{fig_tracts}
\end{figure}


The cerebellum in mammals serves to shape motor activities selected in the basal ganglia by ensuring they are precisely timed and well-coordinated. Such activities are initiated by the basal ganglia with executive oversight from the prefrontal cortex. In humans, the cerebellum also supports cognitive functions such as those involved in reading~\cite{TravisetalHBM-15}. Figure~{\urlh{#fig_White_Matter_Tracts_Long_Distance}{\ref{fig_tracts}}} ({\colorred{B}}) shows the white matter tracts connecting the cerebellum and prefrontal cortex where such abstract cognitive functions originate. 

A white matter bundle called the {\it{arcuate fasciculus}} {\emdash{}} Figure~{\urlh{#fig_White_Matter_Tracts_Long_Distance}{\ref{fig_tracts}}} ({\colorred{A}}) {\emdash{}} provides reciprocal connections between the frontal cortex and association areas in the parietal and temporal lobes plays a key role in attention and the active maintenance of short-term working memory, including support for language processing in the left hemisphere and visuospatial processing in the right hemisphere~\cite{ChicaetalBSF-18}.

The human brain exhibits structure at many scales, the white matter tracts being but one example. A common pattern involves paths that connect multiple circuits that have their own internal components and connections. At a global scale, processing begins in primary sensory areas, propagates forward through dorsal regions integrating additional sources of information to produce composite representations that are processed in the frontal cortex before returning through ventral regions responsible for motivation and action selection.


\subsection{Reciprocity}


Many of the connections within such paths are reciprocal allowing feedback to adjust behavior and improve prediction. Similar reflective and self-corrective patterns arise within subcortical regions including the hippocampal complex and basal ganglia, e.g., between the dentate gyrus and CA1 in the hippocampus and as layered networks inside individual subcomponents such as the mossy fiber network within the dentate gyrus. Each level solves different problems, offers general insights and provides hints about how one might realize such solutions in artificial systems.



\begin{figure}
  \begin{center} 
    \includegraphics[height=150pt]{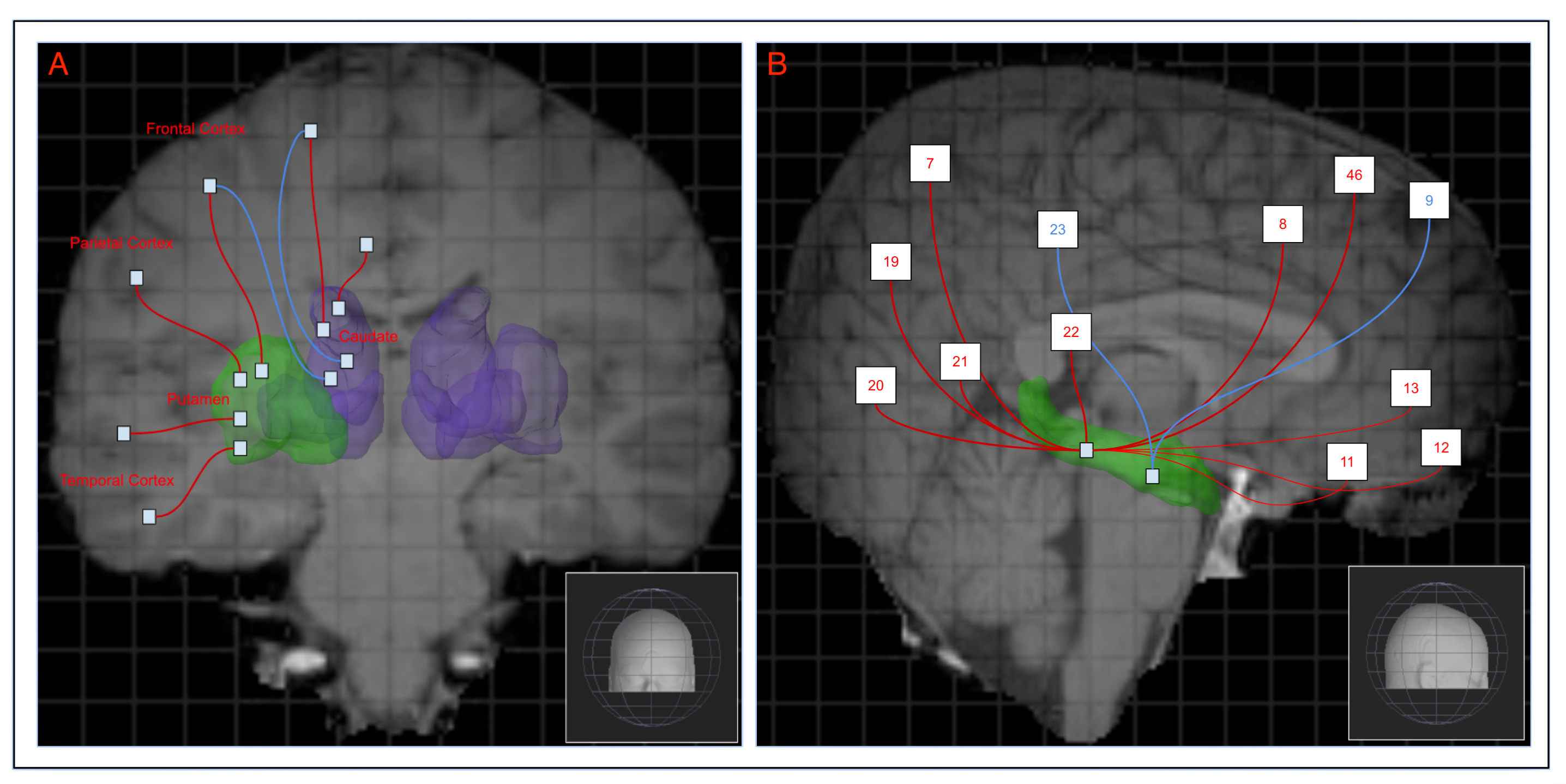} 
  \end{center}
  \caption{The left panel ({\colorred{A}}) illustrates the reciprocal connections between two subnuclei of the basal ganglia, the {\it{putamen}} and {\it{caudate nucleus}}, and locations in prefrontal cortex responsible for influencing action selection. The distinctions between frontal, parietal and temporal cortical areas provide only a very general indication of how their function relates to that of the basal ganglia. The right panel ({\colorred{B}}) highlights reciprocal connections between cortical regions {\emdash{}} identified by the Brodmann areas 7, 8, 9, 11, 12, 13, 19, 20, 21, 22, 23 and 46 {\emdash{}} and the hippocampal complex via the adjacent perirhinal (blue) and the parahippocampal (red) areas. The indicated Brodmann areas generally provide a more nuanced understanding of their possible function than does simply stipulating the cortical lobe they reside in.}
  \label{fig_broadman}
\end{figure}


Figure~{\urlh{#fig_Brodmann_Basal_Ganglia_Hippocampus}{\ref{fig_broadman}}} describes how subcortical nuclei such as the hippocampal complex and basal ganglia interact with cortical regions. Such attributions provide insight on how to construct complex artificial neural architectures composed of simpler subnetworks ostensibly responsible for component functions including perception, action selection and episodic memory.

Here we consider two levels of granularity: the first is coarse grained relying on major anatomical divisions illustrated in Figure~{\urlh{#fig_Human_Brain_Neocortex_Function}{\ref{fig_necortex}}}. The second is somewhat finer grained and relies on areal divisions based on cytoarchitectural distinctions involving cell types, neural processes including dendrites and axons, and other histological characteristics.

The former generally employs Korbinian Brodmann's decomposition of the cortex into 52 {\urlh{https://en.wikipedia.org/wiki/Brodmann_area}{areas}} published in 1909~\cite{Brodmann1909} and revised several times since then to take advantage of more modern staining and imaging technologies as well as improved methods for functional localization. In many cases, identifying the Brodmann area associated with the endpoint of a neural connection can tell us a good deal about the functional relationship between two brain regions.

The left side of Figure~{\urlh{#fig_Brodmann_Basal_Ganglia_Hippocampus}{\ref{fig_broadman}}} ({\colorred{A}}) highlights the reciprocal connections between two subnuclei of the basal ganglia, the {\it{putamen}} and {\it{caudate nucleus}}, and locations in prefrontal cortex responsible for influencing action selection by adjusting input to the basal ganglia and, by way of the thalamus, locations in the parietal and temporal cortex that provide information about the current state relevant to decision making. 

We can often improve functional descriptions if we localize to specific Brodmann areas. For example, the {\it{orbitofrontal cortex}} (OFC) is located in the prefrontal cortex is a region of the frontal lobes involved in the cognitive process of decision-making. In humans it consists of {\it{Brodmann area 10, 11 and 47}}. It is defined as the part of the prefrontal cortex that receives projections from the medial dorsal nucleus of the thalamus, and is thought to represent emotion and reward in decision making~\cite{BotvinickandAnANIPS-09}. The prefrontal cortex, consisting of Brodmann areas 8, 9, 10, 11, 12, 13, 44, 45, 46 and 47, includes the OFC but covers a wider range of functionality.

The right side of Figure~{\urlh{#fig_Brodmann_Basal_Ganglia_Hippocampus}{\ref{fig_broadman}}} ({\colorred{B}}) highlights reciprocal connections between cortical areas {\emdash{}} Brodmann areas 7, 8, 9, 11, 12, 13, 19, 20, 21, 22, 23 and 46 {\emdash{}} and the hippocampal complex via the adjacent {\it{perirhinal}} cortex (shown as blue connections) and the {\it{parahippocampal}} cortex (shown as red connections) that are involved in representing and recognizing objects and environmental scenes.

The anatomy of the brain and patterns of connectivity linking its major functional areas provide a structural account that derives from and informs function. However, functional analyses relating to human cognition require technologies that are able to record neural activity or its correlates aligned with relevant behavioral features. Non-human model systems often employed when invasive technology is required.

On the one hand, optogenetics, two-photon microscopy and conventional electrophysiology are able to record from and modify the electrical activity of tens to thousands of neurons at the resolution of a few microns. While this represents progress, the coverage is inadequate for many studies, and the methods are, for the most part, limited to non-human subjects due to the invasive nature of their practical application~\cite{DombeckandTankCSH-11,BoydenBIOLOGY-11,ZhangetalNATURE-10,YizharetalNEURON-11}.

Conversely, fMRI can used to study awake, behaving humans performing a wide range of cognitive tasks, but relies on signals that are at best indirectly related to neural activity as in the case of blood oxygen levels, and that are currently limited to spatial resolutions on the order of tens of millimeters and temporal resolutions on the order of hundreds of milliseconds~\cite{GoenseetalFiCN-16,GloverPMC-11,BuxtonetalNEUROIMAGING-04}.

Moreover, the electrical activity of individual neurons is but one marker for neural function. Other pathways including diffuse signaling by way of chemical neuromodulation and genetic activity and protein transport at the cellular level are becoming increasingly important as markers for behavior at multiple time scales~\cite{WangandWangFiP-19}. Despite these limitations, neuroscientists have made considerable progress by combining information from different model systems using multiple recording technologies.


\subsection{Sensorimotor Hierarchy}
\label{subsection_sensorimotor_hierarchy}


Much of the cortex is in the business of learning representations of concepts relevant to survival\footnote{%
  The word {\it{perspicacity}} refers to a clarity of perception that enables one to recognize subtle differences between similar physical objects or abstract concepts. It is employed in this context to call attention to the fact that attention, exploration, perception, and prediction are inextricably linked in complex biological systems~\cite{RaoandBallardNATURE-NEUROSCIENCE-99,BarlowNC-89,BaddeleyQJoEP-86,ClarkBBS-13}.}.
Perception is the means by which we apprehend and act on the physical realization of the concepts we have learned. It seems obvious that perception serves action. It may not seem so obvious that action serves perception, but the fact is we are almost always moving our head, hands and torso in order to resolve ambiguities in what we see, feeling the shape of unfamiliar objects in order to grasp them firmly and twisting about to see who is behind us calling our name or to get a better idea of where we've come from in order to ensure we can retrace our steps. These are complex sensorimotor activities we depend on every day.

In thinking about physically realizable concepts we think first about what they look, feel, sound and smell like. The sensory cortex is responsible for constructing a hierarchy of representations to characterize such concepts, not to capture everything we sense, but rather to account for what we need to know about concepts to survive. Reconstructing scenes with photographic realism is not what our sensory systems were designed for. Circuits of the primary sensory cortex feed into the circuits of the (unimodal) association sensory cortex that feed into (multimodal) sensory cortex. All of these representations are abstract and yet patterns of regionalization are remarkably preserved within species~\cite{ChenetalNEURON-18,PortuguesetalNEURON-14,KolsteretalJoN-09}.

Concepts arise in patterns of neural activity that account for what we need to know about them, including how they appear to us so we can recognize them, what affordances they offer for us to make use of them and how we might predict their occurrence in decision making. Many of the concepts that are represented in our brains serve to model the dynamics of physical systems that we interact with every day, such as riding a bike, working with tools, opening doors, negotiating stairs and riding escalators in department stores. Just as important, if not more so, are the social dynamics we deal with at work and school with their constantly shifting personal relationships and status rankings. 

If you are a software engineer designing robot control systems, you might give action much the same scrutiny as perception and build a parallel hierarchy of representations that describes the concepts that relate to movement including navigation, articulation and manipulation ranging from servo-motor commands to strategies for moving furniture, but designing or learning these hierarchies independently is generally a bad idea. In mammals, these two hierarchies are tightly coupled to account for how they depend on one another~\cite{FusterPREFRONTAL-CORTEX-15}.

Indeed, determining what sensory representations to learn depends upon and influences what motor representations to learn and {\it{vice versa}}, where we follow the convention of using the term {\it{motor}} as a catchall term for concepts relating to muscles and movement. As pointed out in the introduction, there is evidence to suggest that circuits occurring early in the ventral visual stream code for object-selective features and exhibit large-scale organization characterized by the high-level properties of animacy and object size~\cite{KonkleandCaramazzaJoN-13,LongetalPNAS-18}.


\begin{figure}
  \begin{center} 
    \includegraphics[width=200pt]{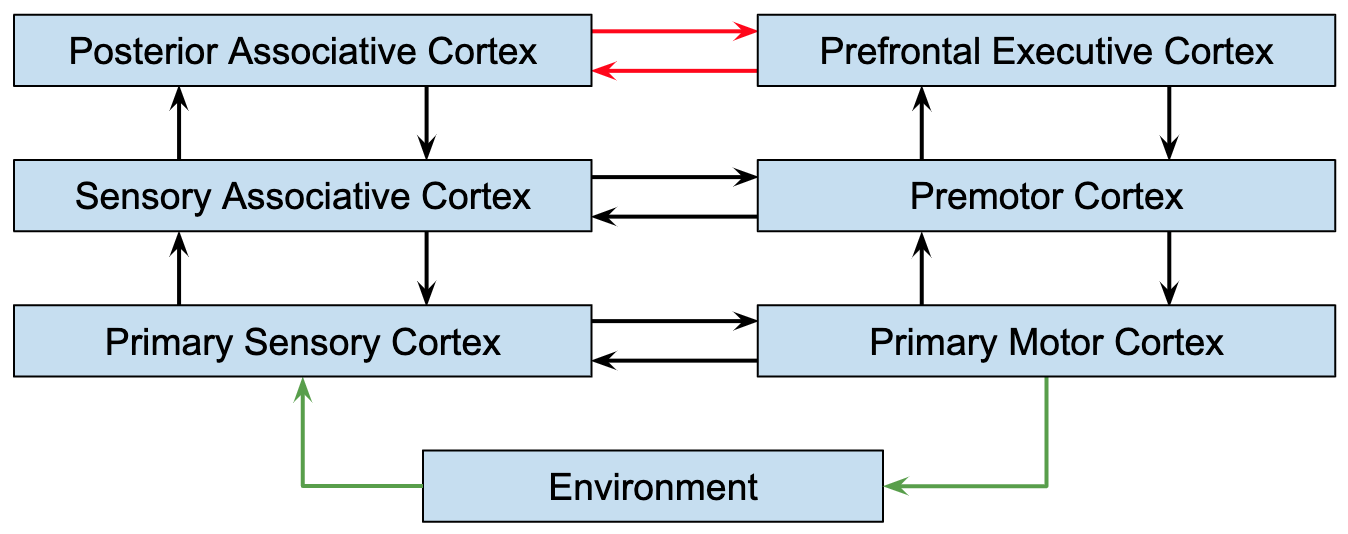} 
  \end{center}
  \caption{A simplified block diagram of the cortex. The column on the left represents the posterior cortex including the occipital, temporal and parietal lobes. The column on the right represents the frontal lobe of the cortex corresponding to the primary motor cortex, premotor cortex (association motor cortex) and prefrontal cortex. Green arrows represent interaction with the environment, black arrows represent sensorimotor abstractions and red arrows indicate cognitive activity relating speech, planning and abstract thinking. See the main text for more detail. Adapted from Figure~8.9 in~\cite{FusterPREFRONTAL-CORTEX-15-CHAPTER_8}}
  \label{fig_coupled}
\end{figure}


Figure~{\urlh{#fig_Coupled_Sensory_Motor_Hierarchy}{\ref{fig_coupled}}} is a simplified block diagram of the cortex organized as two columns. The left column represents the posterior cortex consisting of the occipital, temporal and parietal lobes that are primarily concerned with processing sensory information. The relevant brain areas are summarized in three blocks roughly corresponding to primary sensory cortex, unimodal association cortex and multimodal association cortex stacked so the least abstract concepts are on the bottom and most abstract on the top. The combined area is often referred to as {\it{semantic memory}} and characterized as long-term declarative memory~\cite{BinderandDesaiTiCS-11}. 

The right column represents the frontal lobe of the cortex corresponding to the primary motor cortex, premotor cortex (associative motor cortex) and prefrontal cortex. The primary motor cortex is responsible for creating abstract representations of motor activity throughout the body. The premotor cortex is responsible for integrating sensory and motor abstractions to construct sensorimotor representations. The prefrontal cortex orchestrates cognitive behavior including speech, planning and abstract thinking, and is reciprocally connected to the association areas just mentioned as well subcortical structures including the basal ganglia and hippocampus.

The two columns are connected with one another at multiple levels: by physical interaction with the environment (green arrows), by sensorimotor abstraction and alignment (black arrows), and by cognitive effort in directing activity mediated through subcortical structures (red arrows). This arrangement supports the formation of rich representations that serve a wide range of cognitive function. The sensorimotor connections and feedback through the environment provide an inductive bias to guide learning, ground inference and reduce sample complexity by reducing reliance on labeled data and enabling opportunities for unsupervised learning~\cite{BarlowNC-89}.

Simple as this model of cortical function may seem, it may be one of the most important architectural contributions of neuroscience to the development of artificial intelligence patterned after the human brain. Some of the lessons have already been integrated into the discipline of control theory through exposure to early work in biological cybernetics~\cite{FukushimaBC-80,Lettvinetal59,Jackson1958selected,GibsonPERCEPTION-50,McCullochandPitts43,vonUexk1926theoretical}, but some of the most important lessons impact the application of machine learning in building autonomous embodied systems including robots and digital assistants as alluded to above. 



\subsection{Basal Ganglia}
\label{subsection_basal_ganlia}


There is a long history of neuroscientists constructing computational models of human cognition~\cite{McClelland79,McClellandandRumelhartPR-88,LebiereandAndersonCSS-93,OReillySCIENCE-06,BotvinickPTRS_B-07}. Different modeling tools make different assumptions and support different levels of detail from rule-based systems to spiking neurons~\cite{OReillyetalLEABRA-16,OReillyetalCCN-12,RasmussenetalPLoS-ONE-17,Eliasmith2013,BlouwandEliasmithCSS-13,JilketalJETAI-08}. In this paper, we are primarily concerned with computational models that leverage ideas from neuroscience to develop AI systems for practical problems. In this subsection and the next, we take a closer look at the basal ganglia and hippocampus using models from neuroscience that reveal computational principles we can apply in a wide range of practical problems.



In contrast with the relative simplicity of the neocortical architecture, the {\it{basal ganglia}} consist of specialized subcortical nuclei that are related by their evolved function. In the following, we emphasize and simplify some of those nuclei and ignore others to focus the discussion and simplify the biology. The basal ganglia provide the basis for motor activity controlled by circuits in the brainstem and conserved throughout vertebrate evolution for nearly half a billion years. The cerebral cortex has been around in the form of a six-layer sheet tiled with a repeating columnar structure since the early mammals came on the scene in the Jurassic period about 200 million years ago. Our lineage separated from mice around 100 million and from macaques and other old world monkeys around 25 million years ago. The modern human neocortex owes much to these earlier evolutionary innovations but is different in ways that make possible our facility with language, complex social organization and sophisticated abstract thinking. Compared with the basal ganglia, the neocortex is structurally elegant and functionally general.

The basal ganglia have evolved along with our neocortex to provide us with a powerful thinking machine, while at the same time leaving us to make do with some less-than-ideal adaptations. We can simulate a conventional computer in our heads but are limited to fewer than a dozen memory registers. Most of us can't perform long division in our heads even though we might know the algorithm and aided with paper and pencil carry out the necessary computations to produce an answer. We rely on the same basic cognitive machinery we use to list a few names in alphabetical order to perform all sorts of more complicated cognitive tasks. Even simpler, however, is the basic operation of choosing one of several actions to perform next. The basal ganglia play a key role in supporting action selection and it is worth looking at in a little more detail in order get a handle on some of the parts of the brain that figure prominently in human cognition. Recall that the cortex is a sheet of neural tissue more or less homogeneous in terms of its local structure quite unlike almost any other part of the brain except for the cerebellar cortex. The cortex sits on top of a structure called the {\it{thalamus}} which among other things serves as a relay in passing information back and forth between the cortex and various subcortical nuclei.

The basal ganglia consist of a bunch of circuits, of varied size, sometimes but not always consisting primarily of one cell type, sometimes but not necessarily compactly clustered together, sporting projections that seem to wander off aimlessly, but more or less located above the brainstem and below the cortex. As a general principle, if a signal sets off along some path exiting from a circuit, then expect some derivative of that signal to appear later reentering the circuit to serve as feedback. Everything about the brain, and your entire body for that matter, has to be carefully regulated to maintain a dynamic state of equilibrium, and unlike human designs, evolution is generally not able to cleanly separate the parts of the circuit that perform computations in service to behavior from those that deal with respiration, immune response, waste removal, cell repair, death and regeneration, etc. 


\begin{figure}
  \begin{center}
    \includegraphics[width=4.0in]{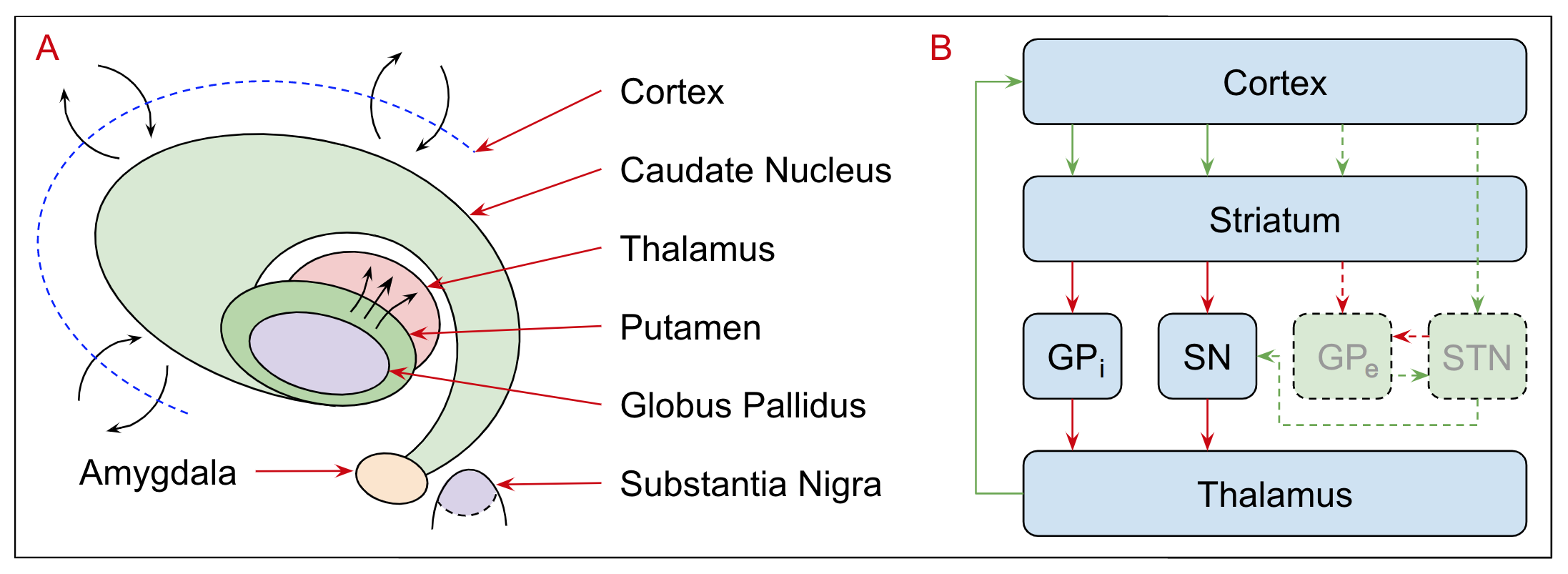} 
  \end{center}
  \caption{The left panel ({\colorred{A}}) provides a highly stylized anatomical drawing of the basal ganglia. Figure~\ref{fig_broadman} ({\colorred{A}}) provides more anatomical detail while the above drawing abstracts from the structural detail in order to simplify the functional account. The block diagram shown in the right panel ({\colorred{B}}) depicts the primary components involved in action selection as functional blocks. The blocks shown in blue represent components in the {\it{direct path}} and are described in the text proper. The blocks shown in light green with dashed borders represent additional components that contribute to the {\it{indirect path}}. Good explanations of the indirect path are described in O'Reilly~\etal{}~\cite{OReillyetalCCN-12} or Wang~\etal{}~\cite{WangetalNATURE-NEUROSCIENCE-18} and we return to the basal ganglia in the next section when we look at the executive role of the prefrontal cortex in modulating behavior.}
  \label{fig_basal}
\end{figure}


The basal ganglia are depicted in Figure~{\urlh{#fig_Basal_Ganglia_Anatomy_and_Physiology}{\ref{fig_basal}}} ({\colorred{A}}) taking some artistic license to keep things simple. The thalamus along with another structure called the {\it{striatum}} provide the interface between the cortex and basal ganglia. The striatum is a combination of a number of smaller nuclei that are anatomically and functionally related; they include the Globus Pallidus (GP), Putamen and Caudate Nucleus and aside from their function as part of the striatum, only the GP will figure prominently in our discussion and only one part of it \emdash{} referred to as the {\it{internal}} GP and identified with the "i" subscript to distinguish it from the {\it{external}} part with "e" subscript.

The other players include the Substantia Nigra (SN) which is at one end of the striatum nestled close to the {\it{amygdala}} which is part of the limbic system involved with memory, decision-making and modulating emotional responses, and the Subthalamic Nucleus (STN). You can think of the cortex as integrating sensory and motor information and making suggestions for what action to take next and the amygdala as supplying information pertaining to the possible emotional consequences of taking different actions to be used as input to action selection. Figure~{\urlh{#fig_Basal_Ganglia_Anatomy_and_Physiology}{\ref{fig_basal}}} ({\colorred{B}}) reconfigures these component nuclei into a smaller number of functionally motivated blocks that control two pathways \emdash{} the {\it{direct pathway}} associated primarily with inhibition and consisting of the internal GP and SN and the {\it{indirect pathway}} playing an excitatory role and consisting of the external GP and STN~\cite{OReillyetalCCN-12,WangetalNATURE-NEUROSCIENCE-18},

The lines connecting the functional blocks shown in Figure~{\urlh{#fig_Basal_Ganglia_Anatomy_and_Physiology}{\ref{fig_basal}}} ({\colorred{B}}) imply neural connectivity, with arrows indicating the direction of influence and colors indicating the valence of the influence, green for excitatory and red for inhibitory. In the action selection cycle, the cortex forwards activations that you can think of as suggestions for what action to take next. These suggestions are propagated through the striatum and forwarded along the direct pathway where two stages of inhibitory neurons initially suppress all of the suggestions and propagate signals back the cortex to activate inhibitory neurons that suppress activity at the source. As this cycle continues, an additional process takes place in the indirect path \emdash{} identified with dashed lines \emdash{} that weighs the advantages and disadvantages of the proposed actions taking in information from throughout the cortex and adjusting the inhibitory bias accordingly.

Eventually, one proposal wins out and all of the others are suppressed allowing a single preferred action to be executed. This cycle of exploring the options for acting and then selecting a single action to execute is constantly repeated during your waking hours. Additional machinery in the thalamus and brain stem regulate whether or not to forward suggestions for acting during sleep when your cortex receives no sensory input and hence any suggestions for acting uninformed by sensory input are ill-advised if not outright dangerous. The above description doesn't begin to convey the complexity of what's going on at the level of individual neurons. Suffice it to say that the usual perfunctory summary consisting of "the winner takes all" doesn't begin to do it justice. The subtleties that arise from the way in which the evidence for and against an action proposal is combined, how ties are broken and deciding when enough evaluation is determined sufficient to make a final choice. 



\subsection{Hippocampal Complex}
\label{subsection_hippocampus}


Figure~{\urlh{#fig_Hippocampus_Anatomy_and_Physiology}{\ref{fig_hippo}}} provides a glimpse of how we construct memories of our experience and subsequently retrieve those memories to support a diverse range of cognitive strategies. In this case, the {\it{hippocampus}} will play a central role as did the basal ganglia in the previous example. In the next section, we explore how the basal ganglia work in concert with the hippocampus to support reinforcement learning. For now, our goal is simply to describe the process whereby we consolidate and then encode experience. In doing so we take the opportunity to talk about the process whereby we retrieve memories, reconstruct a version of that past experience to perform counterfactual inference and imagine possibilities that we have never actually experienced. 


\begin{figure}
  \begin{center}
    \includegraphics[width=4.0in]{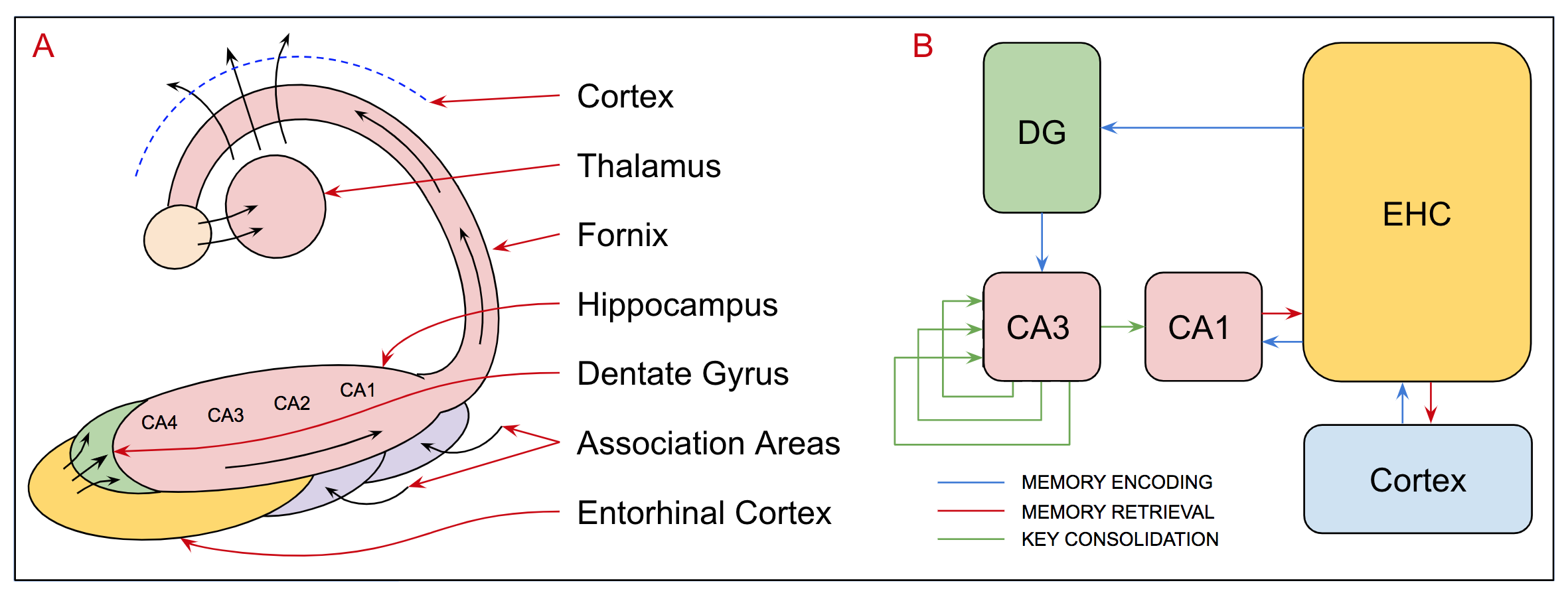} 
  \end{center}
  \caption{On the left you see a cartoon drawing of the hippocampus and related cortical and subcortical areas. The primary components include the entorhinal cortex or EHC, the dentate gyrus or DG and two hippocampal (out of four) nuclei referred to as CA3 and CA1. Figure~\ref{fig_broadman} provides additional anatomical detail regarding the connections between cortical regions and the perirhinal and parahippocampal areas adjacent to the hippocampus. The block diagram on the right summarizes the component circuits, along with their projections and reciprocal connections.}
  \label{fig_hippo}
\end{figure}


The name {\it{hippocampus}} like so many biological terms has obscure origins, generally in Latin or Greek and in this case the latter, relating to its shape that looked like a seahorse to some early anatomists. As shown in Figure~{\urlh{#fig_Hippocampus_Anatomy_and_Physiology}{\ref{fig_hippo}}} ({\colorred{A}}) it is primarily comprised of four subnuclei referred to as CA1, CA2, CA3 and CA4, the first two characters in each abbreviation recalling a previous Latin name, {\it{Cornu Ammanonis}} associated with a ram's horn, apparently preferred by even earlier anatomists. These nuclei are capped by the {\it{dentate gyrus}} (DG) at one end and the {\it{fornix}} at the other\footnote{%
  The hippocampus plays an important role in the transfer of information from short-term memory to long-term memory during encoding and retrieval stages. These stages need not occur successively, but are broadly divided in the neuronal mechanisms they require or even in the hippocampal areas they activate. According to Michael Gazzaniga, "encoding is the processing of incoming information that creates memory traces to be stored." There are two steps to encoding: acquisition and consolidation. During acquisition, stimuli are committed to short term memory. Consolidation is where the hippocampus along with other cortical structures stabilize an object within long term memory. ({\urlh{https://en.wikipedia.org/wiki/Hippocampal_memory_encoding_and_retrieval}{SOURCE}})}. 

The hippocampus consists of two nearly identical structures, one in each hemisphere, connected where the parallel tracts of the fornix come together at the midline of the brain. The hippocampus is tightly coupled with the {\it{entorhinal cortex}} (EHC) that plays an important role in memory, navigation and our perception of time. Information flows from the EHC to the hippocampus by one of two pathways: either through the DG to CA3 or via reciprocal connections to and from CA1. The EHC also has reciprocal connections to many cortical areas. Figure~{\urlh{#fig_Brodmann_Basal_Ganglia_Hippocampus}{\ref{fig_broadman}}} ({\colorred{A}}) provides additional anatomical detail. 

In the process of creating a new memory, the hippocampus receives input from multiple cortical areas relevant to current experience, consolidates this information in a condensed format that will enable subsequent retrieval and stores the resulting encoding in memory. In retrieving an existing memory, The EHC starts with cortical activity, typically from motor and sensory association areas, and uses this information to reconstruct a previous memory by activating cortical areas corresponding to the original memory. Before describing how we think such creative consolidation and subsequent reconstruction works, a word about why this process is beneficial might be in order.

Almost every stage of memory is fraught with opportunities to alter stored representations of prior experience. Reconstruction is a creative process in which we are more often than not forced to fill in some details that we might think we observed at the time but actually didn't. In the formation of new memories, consolidation can only make do with whatever information about the experience we have gleaned from observation and committed to short-term memory. If you don't rehearse what you've stored in short-term memory then it will quickly fade, losing detail and potentially introducing errors of omission and commission.



The basic algorithm carried out by the hippocampus and entorhinal cortex working together is illustrated in Figure~{\urlh{#fig_Hippocampus_Anatomy_and_Physiology}{\ref{fig_hippo}}} ({\colorred{B}}). There are two basic processes that we consider here: encoding new memories and retrieving old memories. Encoding involves collecting information gleaned from diverse neural activity originating in multiple cortical regions and consolidating~\cite{MorrisetalNEURON-06} this information to construct a compact encoding that serves as a key or index that will enable subsequent stable \emdash{} meaning reliably consistent even in the presence of distracting information~\cite{EisenbergetalSCIENCE-03} \emdash{} retrieval and reconstruction\footnote{%
   Memory consolidation is a category of processes that stabilize a memory trace after its initial acquisition. Consolidation is distinguished into two specific processes, synaptic consolidation, which is synonymous with late-phase long-term potentiation and occurs within the first few hours after learning, and systems consolidation, where hippocampus-dependent memories become independent of the hippocampus over a period of weeks to years. Recently, a third process has become the focus of research, reconsolidation, in which previously-consolidated memories can be made labile again through reactivation of the memory trace. ({\urlh{https://en.wikipedia.org/wiki/Memory_consolidation}{SOURCE}})}.

EHC receives input from all cortical regions in a condensed form and the axons of EHC pyramidal neurons project primarily to the DG but also to CA1. DG then projects to CA3 which plays a particularly important role in encoding and retrieving memories. CA3 is thought to behave as an autoassociative memory shown here as a recursive neural network. The crucial property of an autoassociative memory is that it is able to retrieve an item from memory using only a portion of the information associated with that item\footnote{%
  {\it{Autoassociative memories}} are capable of retrieving a piece of data upon presentation of only partial information. {\it{Hopfield networks}} are recurrent artificial neural networks that have been shown to act as an autoassociative memory since they are capable of remembering data by observing a portion of that data. Hopfield networks can be trained with a variety of different learning methods including Hebbian learning which is often summarized as "neurons that fire together wire together". ({\urlh{https://en.wikipedia.org/wiki/Hopfield_network}{SOURCE}})}. 



The hippocampus serves as an index, storing different patterns of cortical activity and allowing us to retrieve our memories using only a fragment of what we can recall. The recurrent connections of CA3 are thought to enable this sort of creative reconstruction and play a role in both encoding and retrieving memories. CA3 then projects to CA1 and from there back to the EHC completing the loop and thereby providing recurrent activity involving a much larger circuit.


There are a couple of details that are worth pointing out here as they demonstrate both the strengths and weaknesses of human episodic memory. The first concerns the issue of retrieving a complete memory given a partial index and the second concerns how to retrieve a memory when that memory is similar to one or more other memories, at least in the sense that their respective indices are similar to one another. In the model described here, the first issue is handled by the autoassociative network.


\begin{figure}
  \begin{center}
    \includegraphics[width=3.25in]{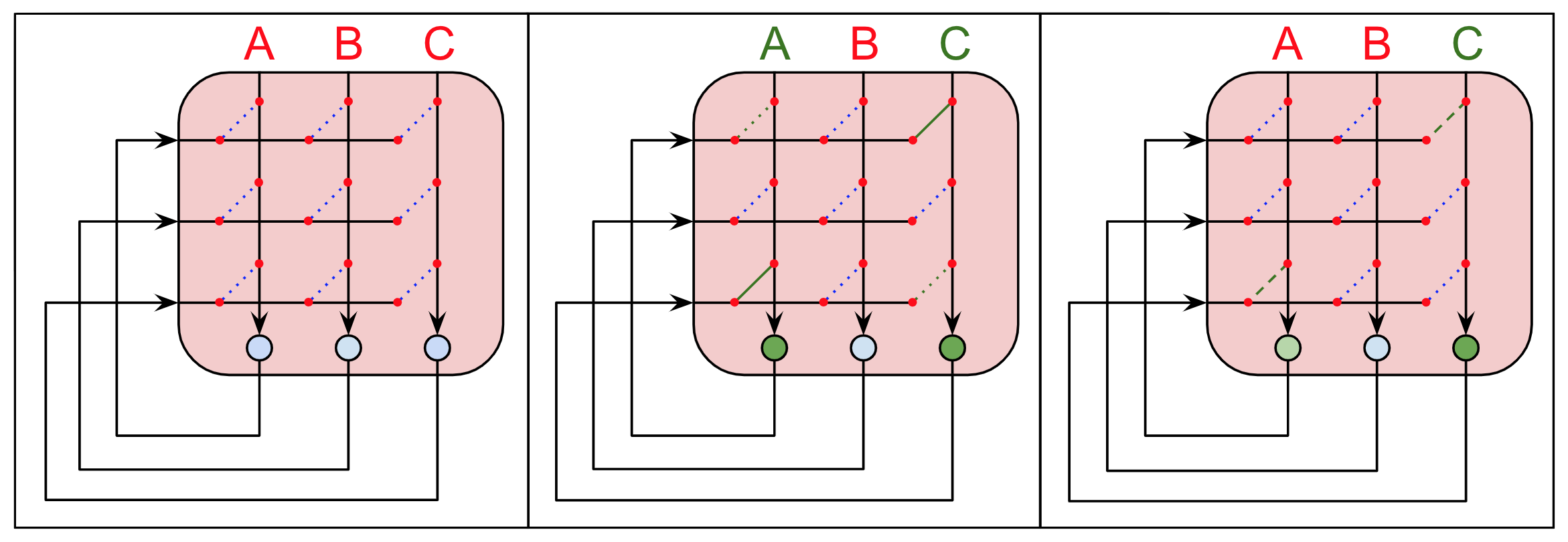} 
  \end{center}
  \caption{The three panels shown here represent the autoassociative network representing the function of CA3 in the hippocampus. Connection weights are shown as diagonal lines, e.g., the dotted blue lines shown in the network on the far left represent the connection weights prior to any training. The middle panel represents the network after encoding the stimulus pattern corresponding to the cortical activation of A and C, and the panel on the far right represents the network, when presented with a partial pattern consisting of just C, employing the recurrent connections of the autoassociator to complete the pattern for the original stimulus and using it to reconstruct the corresponding activation of A and C in the cortex.}
  \label{fig_assoc}
\end{figure}


The triptych shown in Figure~{\urlh{#fig_Hippocampus_Auto_Associative_Network}{\ref{fig_assoc}}} illustrates how the autoassociative network solves this problem. The panel on the far left is meant to indicate the autoassociative network and its initial state. In the middle panel, we assume the input from the dentate gyrus consist of the two sub patterns A and C and illustrates the reciprocal connections that would be strengthened were we to train the network with this composite pattern of activity. CA3 is responsible for encoding these memory specific patterns of activity for all of our memories.  The panel on the far right is intended to illustrate how given a partial pattern, in this case just one of the two representative patterns that comprise the composite pattern shown in the middle panel, is able to reconstruct the other representative pattern by using the trained autoassociative network to first identify and then strengthen the connections in the original composite.

The second detail concerns the possibility that the encodings for two memories are alike enough to be mistaken with one another. A full account of any of the theories explaining how the human brain solves this problem is beyond the scope of what we can go into here but one theory \emdash{} first articulated by David Marr~\cite{MarrandBrindleyPTRS_B-71,WillshawetalPTRS_B-15} \emdash{} posits that, since the dentate gyrus has a larger number of cells than the EHC, its forward projection will tend to produces an expansion recoding in the DG leading to an increase in the separation between the patterns in CA3. 




To complete our account of memory retrieval, we look at how the path that started in the EHC loops back to complete a feedback loop that stabilizes the encoding of memories. So far we've seen how an experience represented by a pattern of activity in the cortex is compressed and represented in the entorhinal cortex which projects this pattern onto the cells in the dentate gyrus thereby increasing the separation between competing patterns the results of which are bound together to generate an index. This index is fed to CA3 where it is incorporated in an autoassociative recursive network so that subsequently when a feature of the original stimulus is present in our conscious experience it activates a subset of the original neurons activated in CA3 and the recurrent connections in the autoassociative network reactivate the remaining neurons completing the pattern that was incorporated when the experience was initially encoded in memory.
  



The remaining step involves explaining how the representation in CA3 reactivates the original stimulus. As shown in Figure~{\urlh{#fig_Hippocampus_Anatomy_and_Physiology}{\ref{fig_hippo}}} ({\colorred{B}}), the entorhinal cortex projects to CA1 in addition to the dentate gyrus. When neurons are projected forward to DG and activated in CA3 they are also activated in CA. Since they are activated at the same time, the connections between the neurons in CA3 and CA1 are strengthened by long-term potentiation\footnote{%
  In neuroscience, long-term potentiation (LTP) is a persistent strengthening of synapses based on recent patterns of activity. These are patterns of synaptic activity that produce a long-lasting increase in signal transmission between two neurons. The opposite of LTP is long-term depression, which produces a long-lasting decrease in synaptic strength. It is one of several phenomena underlying synaptic plasticity, the ability of chemical synapses to change their strength. As memories are thought to be encoded by modification of synaptic strength, LTP is widely considered one of the major cellular mechanisms that underlies learning and memory. ({\urlh{https://en.wikipedia.org/wiki/Long-term_potentiation}{SOURCE}})}.
The result is a stable, sparse, invertible mapping that allows the hippocampus to recreate the original cortical activity patterns during retrieval~\cite{OReillyetalCS-15,McClellandandGoddardHIPPOCAMPUS-97}. Reactivating the same combination of cortical areas as the original stimulus and causing us to reexperience the event as a memory. An additional process called {\it{reconsolidation}} is thought to allow previously-consolidated memories become labile again as a consequence of reactivation\footnote{%
  Memory reconsolidation is the process of previously consolidated memories being recalled and actively consolidated. It is a distinct process that serves to maintain, strengthen and modify memories that are already stored in the long-term memory. Once memories undergo the process of consolidation and become part of long-term memory, they are thought of as stable. However, the retrieval of a memory trace can cause another labile phase that then requires an active process to make the memory stable after retrieval is complete. It is believed that post-retrieval stabilization is different and distinct from consolidation, despite its overlap in function. ({\urlh{https://en.wikipedia.org/wiki/Memory_consolidation}{SOURCE}})}.
See {\urlh{box_patterns}{Box~\colorred{A}}} for detail on storing and retrieving memories in the hippocampus.

   




\begin{center}
  \begin{tcolorbox}[breakable,sharp corners=all,coltitle=black,colbacktitle=white,
    width=\textwidth,boxsep=5pt,left=5pt,right=5pt,
    title={\textbf{Box A: Pattern Separation, Completion and Integration}}]



~~~~As discussed in Section~\ref{subsection_hippocampus}, {\it{pattern separation}} reduces the similarity between input patterns of activity by orthogonalizing inputs to minimize interference between patterns and increase hippocampal storage capacity~\cite{KesnerandRollsNBR-15}. Pattern separation involves primarily DG and CA3 {\emdash{}} see Section~\ref{subsection_hippocampus} for an explanation of the acronyms. The DG maps input from EHC to a much larger and sparsely active GC population. In rats, the number of neurons in the DG exceeds that in EHC by about 5:1~\cite{DrewetalLEARNING-MEMORY-13}. This expansion coding with strong inhibitory interneurons and a competitive learning rule can greatly reduce the overlap between inputs. The DG connects to CA3 mainly through {\it{mossy fibers}} that reliably activate CA3 pyramidal neurons and sustain activation for tens of seconds~\cite{VyletaetalELIFE-16}. Each CA3 neuron receives a small number of these connections from DG so the degree of sparsity is maintained~\cite{KesnerandRollsNBR-15}.

~~~~{\it{Pattern completion}} reconstructs the complete stored pattern given a partial input. Each pyramidal neuron in CA3 receives a large number of synapses from other pyramidal cells forming a recurrent network that serves as an autoassociative memory for pattern completion~\cite{KesnerandRollsNBR-15}. During learning, recurrent connections between active CA3 neurons are strengthened and later when neurons encoding part of an episode are reactivated, they recurrently activate other connected cells to reconstruct the original episode. Basket cells in CA3 form inhibitory synapses to pyramidal cells to dampen excitatory responses thereby emphasizing key features~\cite{NeunuebelandKnierimNEURON-14}. 

~~~~Pattern completion provides access to relevant experience to support decision making in novel situations, and while pattern separation helps downstream discrimination, perfectly orthogonal representations are not ideal in the case we want events that occurred close together to have similar representations. In this case, {\it{pattern integration}} represents related experiences as overlapping populations. There are a number of neural mechanisms suggested to support pattern integration in the hippocampus. We consider two here, the first of which involves {\it{neurogenesis}}. 

~~~~There is evidence that hundreds of new GCs are added to an adult human hippocampus everyday~\cite{SpaldingetalCELL-13}, and stronger evidence suggests that thousands of new GCs are added to rodent’s hippocampus, though not all survive~\cite{KitabatakeetalNCNM-07}. Unlike mature GCs that fire sparsely, immature GCs are more active and have lower threshold for induction of long-term potentiation~\cite{AimoneetalNEURON-09,GeetalNATURE-06,Schmidt-HieberetalNATURE-04}. Aimone~\etal{}~\cite{AimoneetalNEURON-09} posit that a population of hyperactive young GCs could collectively encode events close in time to decrease pattern separation in DG. Others hypothesize that neurogenesis may increase storage capacity by protecting old GCs from new information~\cite{BeckerHIPPOCAMPUS-05,WiskottetalHIPPOCAMPUS-06} or that young active GCs could improve the resolution of memory content~\cite{AimoneetalNEURON-11}. 

~~~~Alternatively, pattern integration might be enabled by recurrent connections involving the hippocampus and neocortex. Recurrent connections in the hippocampus, mainly in CA3 region, can replay an entire episode given a part of it. The replayed episode is backprojected to the neocortex through EHC, that can then recirculate the replayed episode as input to hippocampus to trigger replay of another episode that has overlapping elements with previous one. Kumaran~\etal{}~\cite{KumaranetalTiCS-16} propose that this kind replay between hippocampus and neocortical regions can combine representations of elements that seldom occur together but appear in similar contexts. In addition to integrating experiences with shared elements, backprojection to the medial prefrontal cortex (mPFC) may bias hippocampus to reactivate experiences that are more behaviorally relevant~\cite{SchlichtingandPrestonCOiBS} {\emdash{}} see {\urlh{box_memories}{Box~\colorred{B}}} for more on behavioral relevance. The concurrent presentation of these memories in mPFC may further improve the learning of abstraction relations across episodes.

  \end{tcolorbox}
\end{center}




\section{Architecture}
\label{section_architecture}

The architecture of the human brain, at any scale you choose to consider, bears little or no resemblance to conventional computer architectures. There is no separate program memory, no centralized processing unit, no highly stable, random-access, non-volatile memory and nothing like the digital level of abstraction that enables software engineers to ignore instabilities in the analog circuits that implement logic gates. Since representations (data) are collocated with the transformations (computations) that operate on them and different parts of the brain perform different computations requiring different types of memory, the human brain has to support multiple memory systems.

Human memory\footnote{%
  For students looking for an introduction to memory systems in the brain, you might start with Chapter~24 in~\cite{Bearetal2015} if you own or have access to either the 3rd or 4th edition, or, for a more succinct overview, try Rolls~\cite{RollsARP-01}, Raslau~\etal{}~\cite{RaslauetalAJN-15,RaslauetalAJN-15}, or, most expedient, Wikipedia ({\urlh{https://en.wikipedia.org/wiki/Neuroanatomy_of_memory}{URL}}).} 
is characterized along several dimensions depending on what sort of information is stored, how it is accessed and how long it remains accessible~\cite{CowanPBR-08}. Short-term, long-term and working memory are differentiated on the basis of access, persistence, volatility and the effort required to maintain. Short term is measured in seconds, long term in days, months or years and working memory is essentially short-term memory that can be maintained (with cognitive effort) indefinitely and manipulated (very roughly) analogous to a register in the ALU of a von Neumann machine~\cite{BaddeleyQJoEP-86}.

Declarative memory is defined by the ability to explicitly (consciously) recollect facts, whereas non-declarative memory is accessed unconsciously or implicitly through performance rather than recollection. Episodic memory is generally considered long-term and declarative, and is further differentiated on the basis of the kinds of relationships it can encode, including spatial, temporal and social~\cite{StachenfeldetalNATURE-17,RueckemannandBuffaloNATURE-2017,KumaranandMaguireJoN-05,NealandEichenbaum1993}. Procedural knowledge, including motor, visuospatial and cognitive skills, is encoded in the cerebellum, the putamen and caudate nucleus of the basal ganglia, the motor cortex, and frontal cortex.

To ground the discussion, we introduce the {\it{programmer's apprentice}} as an example of the sort of digital assistants we envision as an application of the technologies presented in this paper. We consider several core components of the apprentice architecture each of which depends on or implements one or more memory systems. Drawing upon concepts covered earlier, we consider three major elements:
\begin{enumerate}
\item the role of the posterior cortex role in supporting declarative knowledge and semantic memory,
\item the basal ganglia and prefrontal cortex as the basis for motivation and executive function, and
\item the hippocampal formation\footnote{%
  The {\urlh{https://en.wikipedia.org/wiki/Hippocampal_formation}{hippocampal formation}} is a compound structure located in the medial temporal cortex that consists of the dentate gyrus, the {\urlh{https://en.wikipedia.org/wiki/Hippocampus}{hippocampus}} proper, subiculum and, depending on whom you consult, presubiculum, parasubiculum and {\urlh{https://en.wikipedia.org/wiki/Entorhinal_cortex}{entorhinal cortex}}.} 
in supporting episodic memory formation, retrieval and consolidation.
\end{enumerate}


\subsection{Embodied Cognition}
\label{subsection_embodied_cognition}


Embodied cognition is the theory that an organism's body shapes its understanding of the environment it inhabits and grounds its perception of and interaction with that environment. Importantly, the environment completes a loop that links perception and action enabling the organism to formulate and test predictive models that guide behavior. Such models serve as the foundation for commonsense reasoning and provide a starting point for understanding a much wider range of concrete and abstract systems, giving rise to a tendency in humans to attribute self-styled agency to both animate and inanimate objects.

To ground our discussion, we consider a personal assistant that works with a software engineer in the role of an apprentice learning on the job, as was common in the guilds and trade associations of medieval cities. The programmer's apprentice we imagine here is a novice programmer but has the intuitive skills of an idiot savant, given that the apprentice has a suite of powerful programming tools as an integral part of its brain. These tools constitute the assistant's body, its peripheral nervous system if you will.

The original programmer's apprentice was the name of project initiated at MIT by Chuck Rich and Dick Waters and Howie Shrobe to build an intelligent assistant that would help a programmer to write, debug and evolve software. Our version of the programmer's apprentice is implemented as an instance of a hierarchical neural network architecture. It has a variety of conventional inputs including speech, text and vision, as well as output modalities including the ability to run code and display program output and execution traces. 

The programmer's apprentice relies on multiple sources of input, including dialogue in the form of text utterances, visual information from an editor buffer shared by the programmer and apprentice and information from a fully {\it{instrumented integrated development environment}} (FIDE) designed for analyzing, writing and debugging code adapted to interface seamlessly with the apprentice as we might move our limbs or direct our gaze. As in case of the legs you were born with, the apprentice has to learn how to use its prosthetic extensions.

\begin{figure}
  \begin{center} 
    \includegraphics[width=200pt]{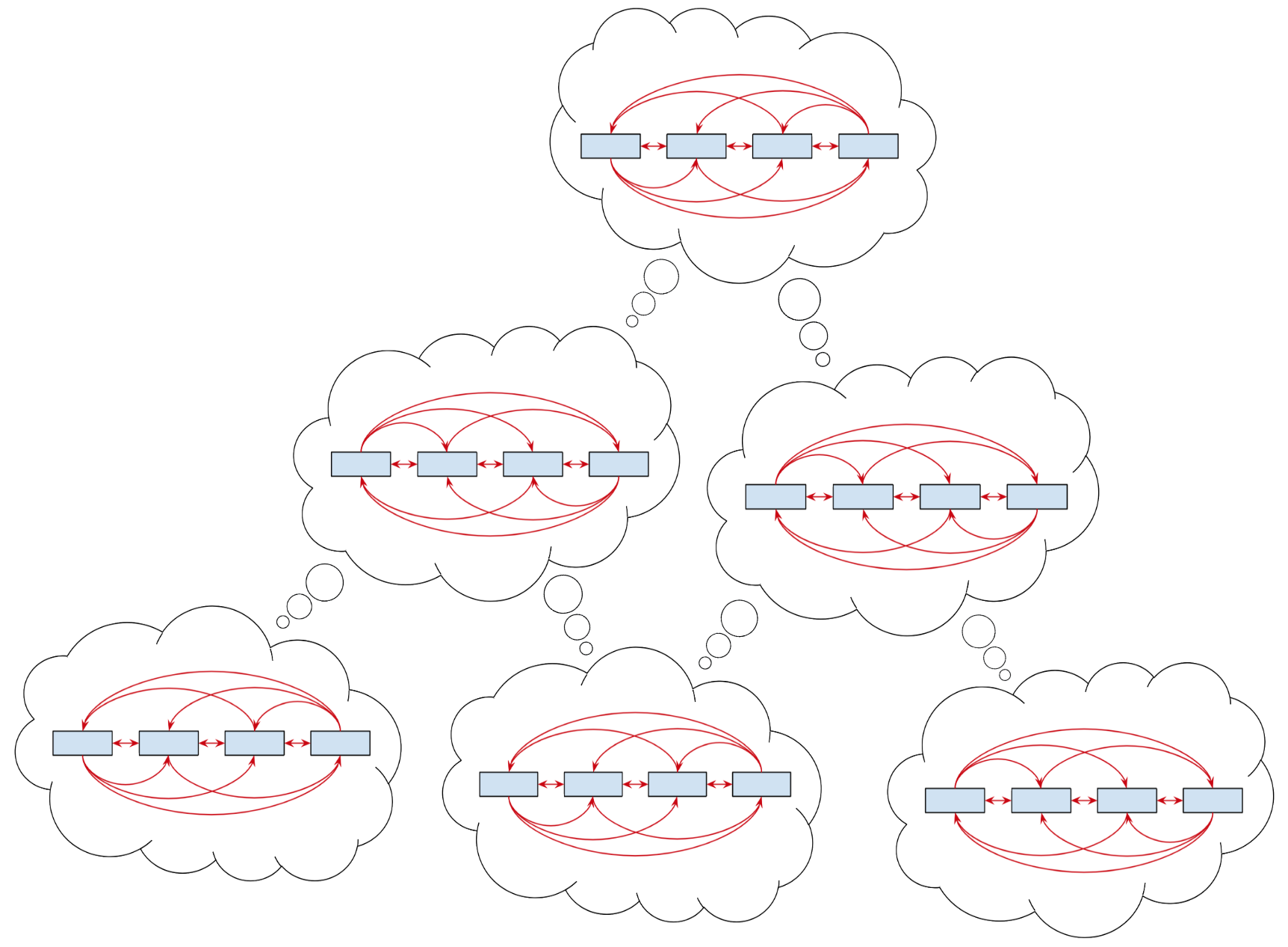} 
  \end{center}
  \caption{The architecture of the apprentice sensory cortex including the layers corresponding to abstract, multi-modal representations handled by the association areas can be realized as a multi-layer hierarchical neural network model consisting of standard neural network components. This graphic depicts these components as encapsulated in thought bubbles of the sort often employed in cartoons to indicate what some cartoon character is thinking. Analogously, the technical term "thought vector" is used to refer to the activation state of the output layer of such a component. All of the bubbles appear to contain networks with exactly the same architecture, where one might expect sensory modality to dictate local architecture. The hierarchical architecture depicted here is modeled after the mammalian neocortex that appears to be tiled with columnar component networks called cortical columns that self-assemble into larger networks and adapt locally to accommodate their input. In practice, it may be necessary to engineer modality-specific networks for the lowest levels of the hierarchy \emdash{} analogous to the primary sensory and motor areas of the neocortex, but more general-purpose networks for the higher levels in the hierarchy \emdash{} analogous to the sensory and motor association areas.}
  \label{fig_posterior}
\end{figure}

This input is processed by a collection of neural networks modeled after the primary sensory areas in the primate brain. The outputs of these networks feed into a hierarchy of additional networks corresponding to uni-modal secondary and multi-modal association areas that produce increasingly abstract representations as one ascends the hierarchy as illustrate in Figure~{\urlh{#fig_Posterior_Cortex_Semantic_Cloud_Memory}{\ref{fig_posterior}}}.
  

Architecturally, the apprentice's FIDE is an instance of a differentiable neural computer (DNC) introduced by Alex Graves and his colleagues at DeepMind~\cite{GravesetalNATURE-16}. The assistant combined with its FIDE corresponds to a neural network that can read from and write to an external memory matrix, combining the characteristics of a random-access memory and set of memory-mapped device drivers and programmable interrupt controllers. The interface supports a fixed number of commands and channels that provide feedback.

The integrated development environment and its associated software engineering tools constitute an extension of the apprentice’s capabilities in much the same way that a piano or violin extends a musician. The extension becomes an integral part of the person possessing it and over time their brain creates a topographic map that facilitates interacting with the extension. We expect the same to occur in the case of the assistant. 


\subsection{Conscious Attention}
\label{subsection_conscious_attention}


Stanislas Dehaene and his colleagues have developed a computational model of consciousness that provides a practical framework for thinking about consciousness that is sufficiently detailed for much of what an engineer might care about in designing digital assistants~\cite{Dehaene2014}. Dehaene's work extends the {\it{Global Workspace}} Theory of Bernard Baars~\cite{Baars1988}. Dehaene's version of the theory combined with Yoshua Bengio's concept of a {\it{consciousness prior}} and deep reinforcement learning~\cite{MnihetalCoRR-13,NairetalCoRR-15} suggest a model for constructing and maintaining the cognitive states that arise and persist during complex problem solving~\cite{BengioCoRR-17}.

Global Workspace Theory accounts for both conscious and unconscious thought with the primary distinction for our purpose being that the former has been selected for attention and the latter has not been so selected. Sensory data arrives at the periphery of the organism. The data is initially processed in the primary sensory areas located in posterior cortex, propagates forward and is further processed in increasingly-abstract multimodal association areas. Even as information flows forward toward the front of the brain, the results of abstract computations performed in the association areas are fed back toward the primary sensory cortex. This basic pattern of activity is common in all mammals.

\begin{figure}
  \begin{center} 
    \includegraphics[height=130pt]{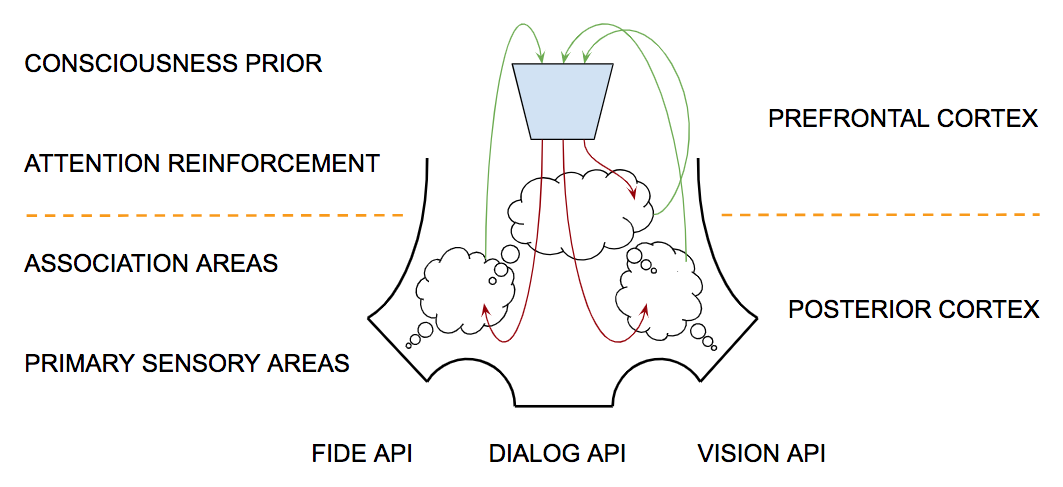} 
  \end{center}
  \caption{The basic capabilities required to support conscious awareness can be realized in a relatively simple computational architecture that represents the apprentice's global workspace and incorporates a model of attention that surveys activity throughout somatosensory and motor cortex, identifies the activity relevant to the current focus of attention and then maintains this state of activity so that it can readily be utilized in problem solving.  In the case of the apprentice, new information is ingested into the model at the system interface, including dialog in the form of text, visual information in the form of editor screen images, and a collection of programming-related signals originating from a suite of software development tools. 
Single-modality sensory information feeds into multimodal association areas to create rich abstract representations. Attentional networks in the prefrontal cortex take as input activations occurring throughout the posterior cortex. These networks are trained by reinforcement learning to identify areas of activity worth attending to and the learned policy selects a set of these areas to attend to and sustain. This attentional process is guided by a prior that prefers low-dimensional thought vectors corresponding to statements about the world that are either true, highly probable or very useful for making decisions. Humans can sustain only a few such activations at a time. The apprentice need not be so constrained.}
  \label{fig_conscious}
\end{figure}

Humans have a large frontal cortex that includes machinery responsible for conscious awareness and that depends on an extensive network of specialized neurons called {\it{spindle cells}} that span a large portion of the posterior cortex allowing circuits in the frontal cortex to sense relevant activity throughout this area and then manage this activity by creating and maintaining the persistent state vectors that are necessary when generating extended narratives or working on complex problems that require juggling many component concepts at once. Figure~{\urlh{#fig_Global_Workspace_Conscious_Attention}{\ref{fig_conscious}}} suggests a neural architecture combining the idea of a global workspace with that of an attentional system for identifying relevant input.


These attentional networks are connected to regions throughout the cortex and are trained via reinforcement learning to recognize events worth attending to according to the learned value function. Using extensive networks of connections {\emdash{}} both incoming and outgoing, attentional networks are able to create a composite representation of the current situation that can serve a wide range of executive cognitive functions, including decision making and imagining possible futures. The basic idea of a neural network trained to attend to relevant parts of the input is key to a number of the systems that we'll be looking at.




\begin{figure}
  \begin{center} 
    \includegraphics[height=150pt]{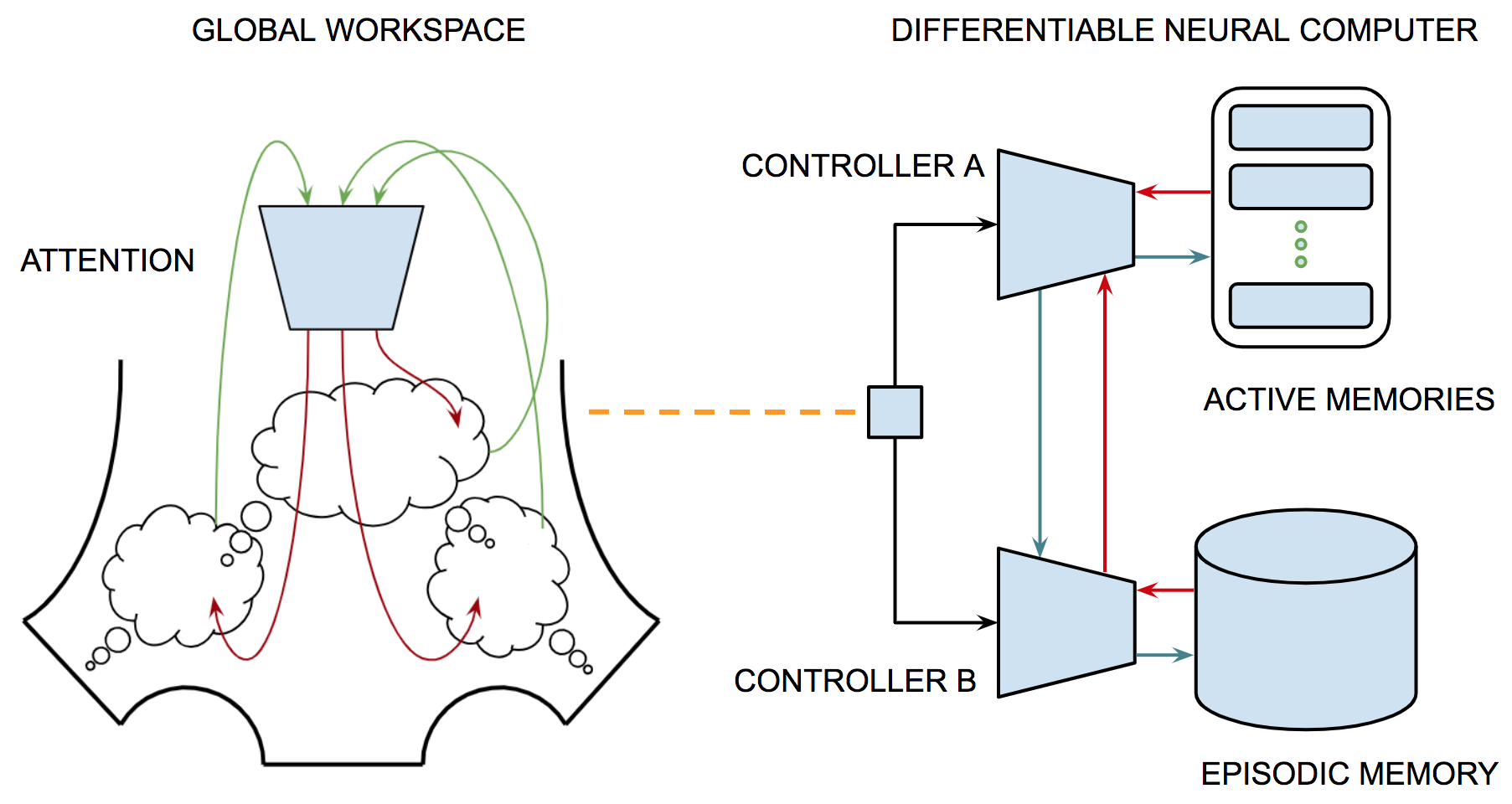} 
  \end{center}
  \caption{You can think of the episodic memory encoded in the hippocampus and entorhinal cortex as RAM and the actively maintained memories in the prefrontal cortex as the contents of registers in a conventional von Neumann architecture. Since the activated memories have different temporal characteristics and functional relationships with the contents of the global workspace, we implement them as two separate NTM memory systems each with its own special-purpose controller. Actively maintained information highlighted in the global workspace is used to generate keys for retrieving relevant memories that augment the highlighted activations. While the associative keys required to access locations only partially match locations, they can can still be used to guide attention allowing the NTM to recognize and even partially merge related locations. In general, locations in memory correspond to thought vectors that can be composed with other thought vectors to shape the global context for interpretation.}
  \label{fig_episodic}
\end{figure}


In their paper~\cite{GravesetalNATURE-16} in {\it{Nature}}, The authors note that "there are interesting parallels between the memory mechanisms of a DNC and the functional capabilities of the mammalian hippocampus. DNC memory modification is fast and can be one-shot, resembling the associative long-term potentiation of hippocampal CA3 and CA1 synapses. The hippocampal dentate gyrus, a region known to support neurogenesis, has been proposed to increase representational sparsity, thereby enhancing memory capacity: usage-based memory allocation and sparse weightings may provide similar facilities." See the discussion of neurogenesis as an algorithmic technique in {\urlh{box_patterns}{Box~\colorred{A}}}.

The global workspace summarizes recent experience in terms of sensory input, its integration, abstraction and inferred relevance to the context in which the underlying information was acquired. To exploit the knowledge encapsulated in such experience, the apprentice must identify and make available relevant experience. The apprentice's experience is encoded as tuples in an NTM that supports associative recall. We'll ignore the details of the encoding process to focus on how episodic memory is organized, searched and applied to solving problems.



\subsection{Action Selection}
\label{subsection_action_selection}




In both neuroscience and artifial intelligence, reinforcement learning problems are typically modeled as Markov decision problems (MDPs). While MDPs can be solved in polynomial time, the size of the state space is often prohibitively large, making practical solution intractable~\cite{LittmanetalUAI-95}. Hierarchical reinforcement learning offers a means of reducing the computational burden by decomposing the state space resulting in a relatively small number of tractable MDPs each of which can be solved independently~\cite{KaelblingICML-93,DietterichJAIR-00,HengstEMLDM-17}. However, the problem of finding an optimal decomposition is itself intractable and hence it is necessary to resort heuristic methods and approximate solutions.

There exist a number of approaches that develop solutions to the problem of hierarchical reinforcement learning (HRL) employing various decomposition strategies, several of which we draw inspiration from~\cite{NarasimhanetalJAIR-18,AndreasetalICML-17,SahnietalCoRR-17,KulkarnietalNIPS-16,BakkerandSchmidhuberIAS-04,MoffaertandNoweJMLR-14,PashevichetalCoRR-18} including a few that relate to biological or biologically plausible models~\cite{RasmussenetalPLoS-ONE-17,DiuketalCRMHOB-13,FrankandBadreCEREBRAL-CORTEX-12,RibasFernandesNEURON-11}. It's important to keep in mind that we are dealing a partially-observable, high-dimensional, continuous state space, and an action space that includes abstract cognitive activities in addition to concrete physical activities that engage the motor system in interacting with the environment. 

In the treatment here, we emphasize the problem of life-long learning as it relates to the nonstationarity of underlying process as a consequence of changes in the external environment and changes in the goals of the agent and the neural substrate available for computation during development and extending on into adulthood. In the case of a growing infant, the changes involve the appearance and maturation of critical circuits and the limitations of finite memory. In both human and machine, internal representations progress from concrete to abstract, building on a foundation grounded in the environment. This maturation in cognitive capability is accelerated by a curriculum that takes advantage of dependencies between concepts. 


\begin{figure}
  \begin{center} 
    \includegraphics[width=350pt]{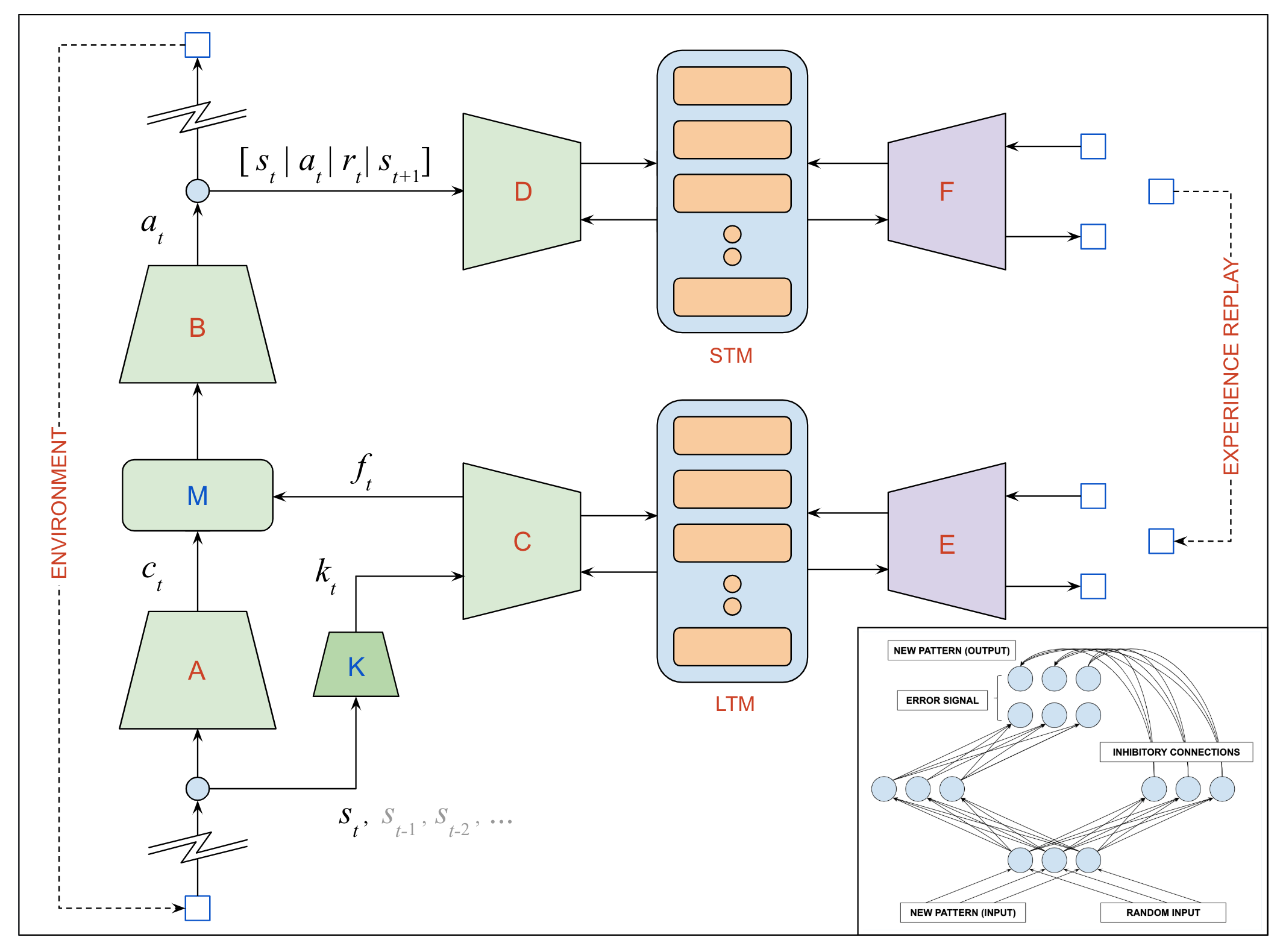} 
  \end{center}
  \caption{The network shown here takes as input a pattern of activation originating in the temporal and parietal lobes and selects an action to perform. The subnetworks labeled \colorred{A} and \colorred{B} are relatively straightforward multilayer neural networks that compute features and generate representations as their output. Network \colorred{A} takes as input a representation of the current state, and generates a representation of the context for action selection. Network \colorblu{K} is an embedding network that takes as input a sequence of states corresponding to recent activity and generates as output a unique key associated with a subspace of the full MDP state space that includes the current state. The box labeled \colorblu{M} corresponds to a location in working memory. The networks \colorred{C}, \colorred{D}, \colorred{E} and \colorred{F} are controllers for two differentiable neural computer (DNC) peripherals that provide storage and access for short-term and long-term memory respectively. The long-term memory is used to store the weights for networks that encode architecturally identical networks {\emdash{}} only the weights are different {\emdash{}} providing specialized expertise in restricted domains corresponding to subspaces of the full MDP state space. The model operates in two modes. In each cycle during the {\it{online mode}}, the \colorred{C} controller loads the selected expert network into location \colorblu{M} where it is fed the output of \colorred{A} and produces the input to \colorred{B}. In this mode, the short-term memory is used to record activity traces that are subsequently used in the {\it{offline}} mode to update the networks stored in long-term memory. The network in the lower-right inset implements a version of {\it{pseudo rehearsal}} as a means of mitigating catastrophic forgetting~\cite{FrenchTiCS-99}.}
  \label{fig_hippo}
\end{figure}


The network model shown in Figure~{\urlh{#fig_Hippocampus_Inspired_Learning_Redux}{\ref{fig_hippo}}} illustrates a system that takes as input a pattern of neural activity originating in the medial temporal and inferior parietal cortex and selects an action to perform. This particular example is meant to illustrate how episodic memory might play an expanded role in action selection. For illustration, patterns of activity serve as proxies for the state of the external environment and are represented in the figure as a sequence $s_t, s_{t-1}, s_{t-2}, ...$. The subnetworks labeled \colorred{A} and \colorred{B} are relatively straightforward multilayer neural networks that compute features and generate representations as their output. Network \colorblu{A} takes as input a representation of the current state, and generates a representation of the {\it{context}} for action selection.


We'll explain the function of the box labeled \colorblu{M} in a moment; assume for now that it generates a representation of the options available for acting in the current context. Network \colorred{B} then takes these suggestions as input and produces as output a representation of the selected action. The boxes labeled \colorred{C}, \colorred{D}, \colorred{E} and \colorred{F} are controllers for two differentiable neural computer (DNC) units that provide storage and access for short-term and long-term memory respectively. The controllers on the left are part of the online system for selecting actions. The controllers on the right are responsible for off-line training during which the recorded actions, along with their associated states and rewards are consolidated in long-term memory using experience replay.

The blue boxes represent stored information in the form of key-value pairs. Each key is associated with a subset or {\it{subspace}} of the set of all states that represents a restricted domain of expertise for selecting actions. The value for this key is a function implemented as a network trained as an expert for the associated subspace. \colorblu{K} is an embedding network that takes as input a sequence of states corresponding to recent activity and generates as output a unique key associated with a subspace of the current state. A given state can belong to more than one subspace and the particular key selected at any given point in time depends on the current state and the immediately previous states in a fixed window. The order of the states matters. 

In the online phase, the embedding network retrieves this key which it forwards to the controller labeled \colorred{C} that uses it to retrieve the expert for the relevant subspace. The box labeled \colorblu{M} corresponds to a location in working memory and in each online cycle the \colorred{C} controller loads the expert subsystem in location \colorblu{M} of working memory where it can be utilized to compute a set of options appropriate for the current state. During off-line periods the system uses the recorded sequences of activity to run some variant of experience replay to update the relevant expert subsystems stored in long term memory~\cite{AndrychowiczetalCoRR-17,SchauletalCoRR-15,LinML-92}.

The training that occurs offline involves adjusting the weights of networks using relatively small samples and so runs the risk of catastrophic interference in transfer learning~\cite{McClellandetalPR-95}. One way in which we hope to ameliorate the adverse consequences of catastrophic interference is by defining separate networks for separate subspaces. The embedding space method mentioned in describing \colorblu{K} is designed to isolate expertise by identifying states that tend to occur together. The hope is that the actions exercised is such states will tend to be interrelated and hence they should be represented using the same network to facilitate their coordination. 

Of course temporal proximity in their occurrence doesn't guarantee they serve the same task since we are always getting distracted or interrupted requiring us to interleave tasks that have very little to do with one another. It may be possible to segment activity streams into coherent tasks in a similar way to how we segment conversations involving multiple speakers~\cite{SeldinetalICSS-01,SeldinetalICML-01}. Alternatively, there has been some success with the method of {\it{pseudo rehearsal}} which consists of retraining existing networks by interleaving new examples with synthetic-examples produced by randomly activating the existing network~\cite{ZhiyuanandBingLML-18,KirkpatricketalCoRR-16,AnsetalCSS-02,FrenchTiCS-99,FrenchCONNECTION-SCIENCE-97,RobinsCONNECTION-SCIENCE-95}.

In this model the STM roughly corresponds to the hippocampus as the storage system for episodic memory. The LTM resembles the cerebellum in the way that it essentially compiles prior activity to construct a set of programs each of which spans some portion of the overall state space. As described above, the STM is only used for temporary storage awaiting off-line replay to consolidate recent memories. An alternative is to maintain a much larger collection of episodic memories that can be used in a manner similar to that suggested in Gershman and Daw who posit that we routinely draw upon our stored memories in the hippocampus to figure out what to do in novel situations not covered by our other sources of procedural knowledge~\cite{GershmanandDawANNUAL-REVIEWS-17}. See {\urlh{box_memories}{Box~\colorred{B}}} for more detail concerning episodic memory and experience replay.



\begin{center}
  \begin{tcolorbox}[breakable,sharp corners=all,coltitle=black,colbacktitle=white,
    width=\textwidth,boxsep=5pt,left=5pt,right=5pt,
    title={\textbf{Box B: Replaying Experience, Consolidating Memory}}]

    
~~~~When we encounter a new experience in the environment, we do not act independently of the past, but rather, our past experiences substantially inform our present decisions. Here, we introduce the basic principles of \textit{hippocampal replay} and a few key ways in which it has motivated reinforcement learning algorithms.

~~~~Replay is the process by which hippocampal representations of previous experiences are sequentially reactivated~\cite{CarretalNATURE-NEUROSCIENCE-11}. Studies show that cells in the rodent hippocampus replay past experiences to stabilize behaviorally relevant memories~\cite{OlafsdottirCURRENT-BIOLOGY-18,JooandFrankNATURE-REVIEWS-NEUROSCIENCE-18}. Though initially observed in spatial tasks, recent work suggests that non-spatial task states are also replayed, and that this phenomenon is common in humans~\cite{SchuckandNivDOI-18}. 

~~~~In the reinforcement learning literature, the \textit{experience replay} algorithm was introduced as an analogical framework in online learning agents~\cite{LinML-92}. Transitions containing state, action, and reward information are sequentially stored in memory and sampled randomly for learning. Randomly replaying old memories not only allows decorrelation of consecutive experiences encountered during data collection, but also enables reuse of training data, increasing sample efficiency, and encourages resampling of rare experiences, potentially alleviating forgetting. 

~~~~A relatively well-studied question is \textit{what to replay}. Some studies suggest the correlation of replay frequency with \textit{novelty} approximated by temporal difference (TD) error~\cite{FosterandWilsonNATURE-06}, and others with high \textit{reward}~\cite{OlafsdottirCURRENT-BIOLOGY-18}. In particular, dopaminergic release, which encodes both novelty and reward~\cite{MenegasetalELIFE-17}, enhances {\textit{sharp wave-ripple}} activation \emdash{} the basic unit of replay. Yet other studies show that experiences more \textit{vulnerable to forgetting} are more likely to be replayed~\cite{SchapiroetalNATURE-COMMUNICATIONS-18}. While the exact selection algorithm is unknown, the observed association with novelty inspired the \textit{prioritized experience replay} algorithm which samples experiences with probabilities weighted by their TD errors and is now consistently preferred to the originally proposed uniform sampling variant~\cite{SchauletalCoRR-15}. 

~~~~The significantly less studied question is \textit{what happens during replay}. Besides re-learning of experiences, many neuroscientists support the idea that replay also serves as a substrate for \textit{memory consolidation} \emdash{} the gradual integration of new experiences processed into existing knowledge representations in the neocortex~\cite{WilsonandMcNaughtonSCIENCE-94,McClellandetalPR-95,KarlssonandFrankNATURE-NEUROSCIENCE-09,BendorandWilsonNATURE-NEUROSCIENCE-12,KumaranetalTiCS-16}, as to stabilize memories against interference. The idea is that replaying information stored in memory will encourage synaptic consolidation processes. 

~~~~While we lack a precise understanding of the underlying mechanisms of consolidation in the brain, in our architecture we frame consolidation as the process by which experiences are used to update expert subsystems stored in long-term memory. We propose an adaptive replay algorithm whereby experiences with contexts similar to the current context are replayed and thus preferentially consolidated into long-term storage. Since action selection directly depends on the relevant expert network drawn from long-term memory, we can ensure to maximally update the currently relevant expert network with existing memories related to its corresponding context. This algorithm is partly inspired by the result by~\cite{JooandFrankNATURE-REVIEWS-NEUROSCIENCE-18} whereby they observed that when a rat pauses at a branching point in a maze, it replays representations of trajectories in the past with similar context to drive its present decision-making. 

~~~~There exist many other cognitively inspired variants of experience replay. One example is {\textit{hindsight experience replay}}~\cite{AndrychowiczetalCoRR-17}, where the agent pretends that whatever state it reaches had been the goal state from the start and learns from the experience regardless of whether it actually succeeded, just as humans can learn from undesirable outcomes. 



  \end{tcolorbox}
\end{center} 



In this case, the LTM stores what can be thought of as subroutines or libraries for solving routine problems. Used in the manner described in Gershman and Daw~\cite{GershmanandDawANNUAL-REVIEWS-17}, the DNC labeled STM more closely captures the functionality of the hippocampus in combining short-term and long-term episodic memories with specific procedural knowledge based on past experience that may or may not be common enough to warrant compiling as a standalone library. The dentate gyrus is best known for its ability to separate patterns to avoid mistaking one pattern for another. Less well understood is a possible complementary role that involves integrating similar patterns.

The ability to draw upon episodic memory to select what to do in situations similar to those encountered in the past provides a simple form of one-shot learning. It could enable us to make predictions, perform hypothetical reasoning and put ourselves in someone else's shoes assuming that our ability to retrieve memories allows us match situations that we find ourselves that we haven't experienced, but know from someone else's experience. It might avoid some of the problems with interference if the process of integrating new procedural knowledge with old could be spread out over longer periods if, say, each time you encounter a similar situation you make only minor adjustments to the weights of the associated subspace expert network. 

\subsection{Executive Control}
\label{subsection_executive_control}


There is a growing consensus and a fair bit of evidence to support the hypothesis that the human frontal cortex is in charge of executive control, goal-directed planning and abstract thinking. There are differences in opinion about how these cognitive processes are implemented and how they coordinate their activities with that of the rest of the brain. One thing that seems clear is that the frontal cortex and in particular the prefrontal cortex employs many of the same strategies as do networks elsewhere in the brain, both cortical and subcortical.

In particular, circuits in the prefrontal cortex recapitulate the coarse-to-fine, concrete-to-abstract feature hierarchies that we see in the sensory, motor and somatosensory cortex. They exhibit the profuse reciprocal recurrent connections between levels of abstraction that enable us to generalize on the basis of relatively small amounts of information, learn to make accurate predictions in an unsupervised manner depending on observations and interactions with the environment to ground our conclusions, and that provide the foundation for constructing a rich repertoire of representations that serve decision-making.

The neural correlates of abstract thinking, including the circuits that enable us to solve practical problems as well as pursue pure mathematics, are generally agreed to be located in the prefrontal cortex with reciprocal connections throughout the rest of the cerebral cortex, the cerebellar cortex and subcortical regions including the basal ganglia, hippocampal formation and parts of the limbic system involved with emotion, motivation and episodic memory. See {\urlh{box_abstract}{Box~\colorred{C}}} for more detail regarding abstraction, hierarchy and executive oversight in the prefrontal cortex.



\begin{center}
  \begin{tcolorbox}[breakable,sharp corners=all,coltitle=black,colbacktitle=white,
    width=\textwidth,boxsep=5pt,left=5pt,right=5pt,
    title={\textbf{Box C: Hierarchy, Abstraction and Executive Control}}]

    
~~~~The prefrontal cortex (PFC) is generally considered to be responsible for executive cognitive control and enabling the synthesis of novel behavior. Here, we briefly review prefrontal anatomy, development and physiology, focusing on three key executive cognitive functions: {\it{attentional set}}, {\it{working memory}} and {\it{action selection}}. For each function, we suggest how our current understanding might lead to new architectures and algorithms for AI systems.

~~~~The PFC sits atop a group of hierarchically organized sensory and motor areas in the cortex enforced through reciprocal anatomical connections~\cite{FusterPREFRONTAL-CORTEX-15}. This arrangement, referred to as {\it{Fuster’s hierarchy}}, motivates computational models of the PFC that posit the development of highly abstract representations of the sensorimotor context that can be used understand what we perceive and direct how we act~\cite{BotvinickPTRS_B-07}. In addition to connections between layers, Fuster’s hierarchy stipulates reciprocal connections between sensory and motor areas of cortex at the same level of abstraction within each layer of the hierarchy. This intralayer connectivity between perception and motor suggests that action representations feed back into and enhance perception, a principle codified in the notion of {\it{corollary discharge}}~\cite{mccloskey2011corollary}.

~~~~Beyond the model of network interactions, intralayer connectivity in Fuster's hierarchy suggests that each layer of the hierarchy along with the layers below but excluding those above, forms a self-contained perception-action loop. Evidently the neocortex undergoes a series of developmental stages with the PFC among the last areas to mature~\cite{guillery2005postnatal}. This implies training of a complex agent may need to unfold in a manner akin to greedy layer-wise deep network training~\cite{BengioetalNIPS-07,belilovsky2018greedy}, with developmentally-staged, abstraction-comparable, layer-wise learning of the coupled sensorimotor features.

~~~~{\it{Attentional set}} (ASET) refers to the preparation of downstream perception and motor cortices for expected stimuli or action. ASET is exhibited in {\it{cued-attention tasks}} where, in anticipating a visual stimuli, PFC and V4 will be active before the stimuli is given~\cite{sylvester2009anticipatory}. ASET suggests existing inhibitory attention masks may be augmented with additive excitatory attention, allowing a neural network to reduce bottom-up input needed for neuron stimulation or cause neuron firing in the absence of sensory input altogether. Allowing the controller to generate new patterns via network activation even suggests a new model of imagination, with improvements in both sensory synthesis~\cite{GregoretalCoRR-15} and planning~\cite{PascanuetalCoRR-17}.

~~~~{\it{Working memory}} is the maintenance of recent stimuli for subsequent action planning. Working memory consists of groups of coupled neural circuits in the PFC called {\it{stripes}} that are connected to potential target stimuli in sensory cortex and access controlled by circuits in the basal ganglia. Computational models of working memory~\cite{HazyetalPTRS-07} include implementations similar to the recurrent memory circuit of an LSTM cell~\cite{HochreiterandSchmidhuberNC-97}, more exotic architectures involving stacked LSTMs~\cite{graves2013speech} and multiple memory stripes manipulated by a central controller.

~~~~In {\it{action selection}}, the PFC generates many actions that are approved or denied by the basal ganglia; both basal ganglia and orbitomedial PFC receive dopaminergic afferents originating in midbrain structures, providing a reward signal that reinforces learning~\cite{FusterPREFRONTAL-CORTEX-15}. Computational models of dopaminergic systems~\cite{o2007pvlv} point to an architecture similar to existing actor-critic models~\cite{MnihetalCoRR-16}; a key improvement is the modeling of {\it{reward inhibition}}, whereby learning ceases for repetitive stimuli. Suppression of reward to prevent response overfitting could aid in tackling other problems such as reward hacking~\cite{amodei2016concrete}, catastrophic forgetting, and lifelong learning~\cite{ZhiyuanandBingLML-18}, all challenges in effectively managing the learning process.

  \end{tcolorbox}
\end{center}


With respect to hierarchical goal-based planning, there is growing evidence pointing to a set of adjoining regions in the prefrontal cortex that are responsible for how abstract plans are initially selected, subsequently refined and finally realized as concrete actions. These same regions also appear to be involved in relational reasoning from simple binary relations to higher-order relationships.

These theoretical observations combined with behavioral studies and fMRI recordings have led to a number of computational models of hierarchical planning that exhibit similar patterns of cognitive activity. In particular, cognitive neuroscientists have developed models of how such abstract hierarchical reasoning in the prefrontal cortex is related to what we know about how the basal ganglia and areas of the limbic system involved in motivation contribute to action selection~\cite{OReillySCIENCE-06}. 


\begin{figure}
  \begin{center} 
    \includegraphics[height=150pt]{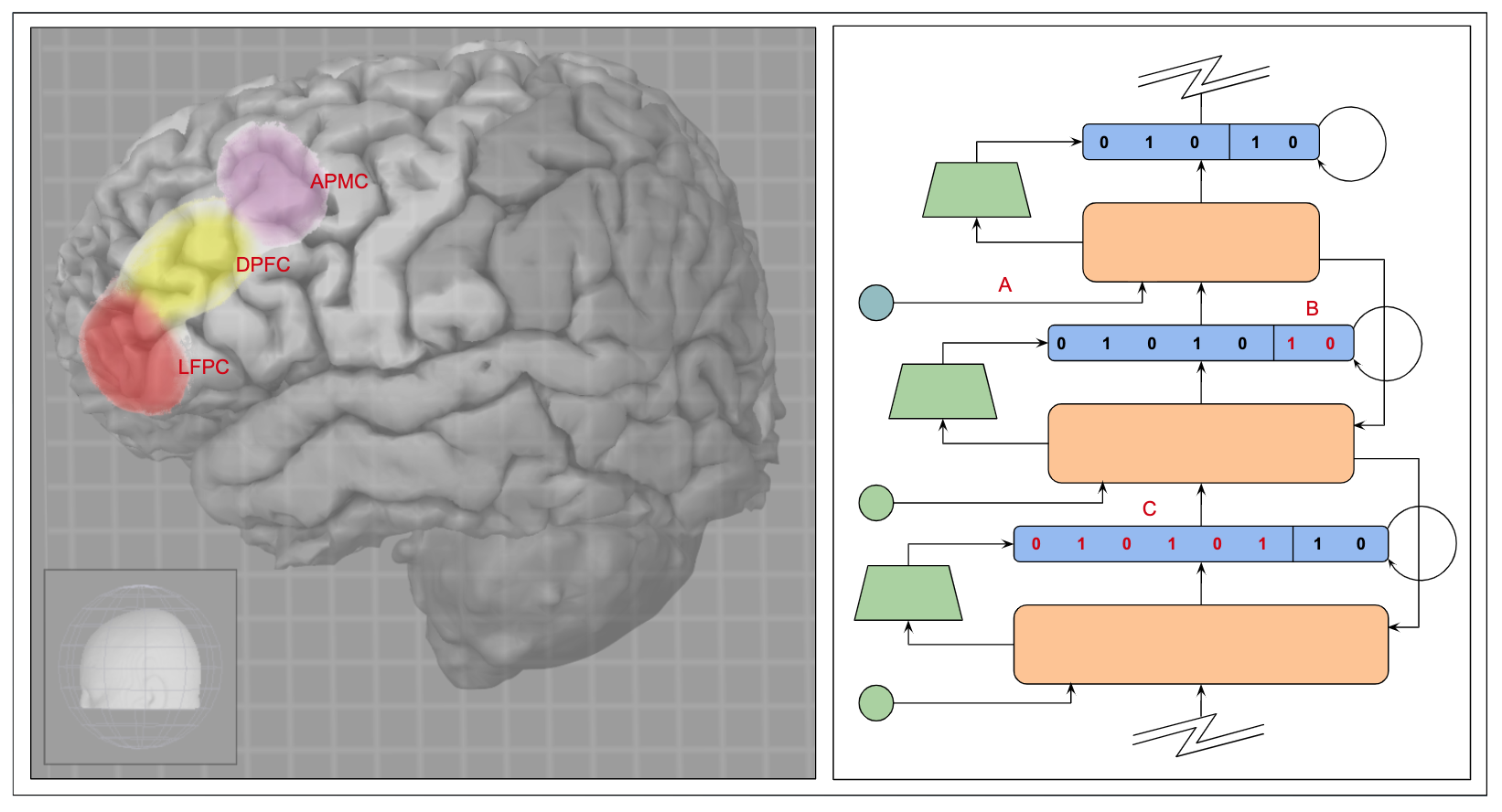} 
  \end{center}
  \caption{%
    The panel on the left highlights three areas of the prefrontal cortex shown in the figure from left to right (rostro-caudal) and referred to in the text as the {\it{lateral frontal polar cortex}} (\colorred{LFPC}) {\it{dorsolateral prefrontal cortex}} (\colorred{DPFC}) and {\it{anterior premotor cortex}} (\colorred{APMC}). According to the theory first articulated by Joaqu\'{i}n Fuster and subsequently refined David Badre~\cite{BadreandWagnerNEURON-04}, Mark D'Esposito~\cite{DEspositoetalNATURE-95} and Etienne Koechlin~\etal~\cite{KoechlinetalSCIENCE-03} and their colleagues, as actions are specified from abstract plans to concrete responses, progressively posterior regions of lateral frontal cortex are responsible for integrating more concrete information over more proximate time intervals. This process of progressive articulation does not correspond to different stages of execution so much as to how actions are selected, maintained and inhibited at multiple levels of abstraction~\cite{BadreTiCS-08}. The panel on the right shows a simple neural-network model of the brain regions aligned with the rostro-caudal axis of the frontal cortex and hypothesized to account for how action representations are selected, maintained and inhibited at multiple levels of abstraction. The neural-network model is described in more detail in the main text, but a few points are in order here: {\colorred{A}} {\emdash{}} different abstraction layers may include input from other sources, e.g., natural language embeddings, that are only required at particular levels of abstraction; {\colorred{B}} {\emdash{}} each recurrent level of the abstraction hierarchy includes state variables encoding information that would typically appear on the call stack in a conventional computer architecture; {\colorred{C}} {\emdash{}} attentional layers mask (suppress) input that is not determined to be relevant to decision making at a given time and level of abstraction resulting in a sparse context vector.}
  \label{fig_prefer}
\end{figure}


The network shown on the right in Figure~{\urlh{#fig_Prefrontal_Hierarchy_Biology_Technology}{\ref{fig_prefer}}} consists of three subnetworks that roughly align with the {\urlh{https://en.wikipedia.org/wiki/Frontal_lobe}{lateral frontal polar cortex}} (bottom), {\urlh{https://en.wikipedia.org/wiki/Dorsolateral_prefrontal_cortex}{dorsolateral prefrontal cortex}} (middle) and {\urlh{https://en.wikipedia.org/wiki/Premotor_cortex}{anterior premotor cortex}} (top) as shown in the figure. Each of the subnetworks is composed of three elements: a recurrent multilayer perceptron constructed of interleaved convolutional and max-pooling layers shown in orange, a multilayer attention network shown in green and a masking layer in blue that selectively suppresses a subset of the outputs of the convolutional stack in accordance with the output of the attention network.

Input to each of the three subnetworks includes areas of associative activity throughout the sensory and motor cortex as well as areas corresponding to higher-level abstractions located in the frontal cortex responsible for abstract thought and subcortical regions responsible for motivation. While not emphasized here, the active maintenance in working memory of information originating from these sources\footnote{%
  Susan Courtney provides an excellent overview of the many sources of information that are utilized by cognitive functions supported in the frontal cortex~\cite{CourtneyCABN-04}. In particular, her articulation of the role of attention and cognitive control aligns with the views that we've emphasized in class and that drive our designs:
\begin{quotation}
  [The circuits in the prefrontal cortex that drive goal directed planning and executive control] receive multimodal information about the current environment and have access to previously stored memories. The prefrontal cortex's extensive outputs allow for direct control of motor behavior, but they may also influence behavior indirectly by altering perceptual and cognitive representations and influencing the storage and re- trieval of long-term memories.\\
  I suggest that attention and cognitive control are not directed actions or specific processes contained within any particular set of brain regions. Rather, what we experience and observe that we call attention and cognitive control are emergent properties dependent on the dis- tributed representation of all types of information, both that available from present perceptual input and the information currently sustained in WM, including contextual and motivational information.
\end{quotation}}
is critical for the cognitive activities that these networks support~\cite{CourtneyCABN-04,Goldman-RakicARN-88}. The outputs are fed to a network (not shown) that serves as the interface for the peripheral motor system (the fully instrumented integrated development environment (FIDE) in the case of the programmer's apprentice) which could play the role of the basal ganglia and cerebellum, but could also be considerably simpler depending on the application.

Figure~{\urlh{#fig_Prefrontal_Hierarchy_Biology_Technology}{\ref{fig_prefer}}} is just a sketch employing familiar neural network components to make the point that building these architectures out of standard components is not the most significant challenge. The real challenge is in training them as part of larger system with lots of moving parts. The expectation here, as in the model sketched in Figure~{\urlh{#fig_Hippocampus_Inspired_Learning_Redux}{\ref{fig_hippo}}}, is that end-to-end training with stochastic gradient descent isn't going to work, and that training will likely require some form of layer-by-layer developmentally-staged curriculum learning~\cite{LampinenetalCoRR-19,GulcehreetalCoRR-16,BengioetalCoRR-15,BengioetalICML-09} and a strategy for holding some weights fixed while adjusting other weights to account for new information and avoid problems like catastrophic forgetting.


\subsection{Digital Assistants}
\label{subsection_digital_assistants}



We focus on automated programming for several reasons: deep neural networks have recently demonstrated progress on automated code synthesis and program repair by leveraging existing technologies; computers and, in particular, modern integrated development environments present a rich alternative to the dominant simulation environments; programming is challenging for humans and machines alike and we foresee opportunities to increase the productivity of software engineers. That said, in this section we emphasize basic tools that enable the assistant to encode, represent and manipulate fully differentiable programs.



The integrated development environment and its associated software engineering tools constitute an extension of the apprentice’s capabilities in much the same way that a piano or violin extends a musician or a prosthetic limb extends someone who has lost an arm or leg. The extension becomes an integral part of the person possessing it and over time their brain creates a topographic map that facilitates interacting with the extension\footnote{%
  In the mammalian brain, information pertaining to sensing and motor control is topographically mapped to reflect the intrinsic structure of that information required for interpretation. This was early recognized in the work of Hubel and Wiesel~\cite{HubelandWieselJoP-68,HubelandWieselJoP-62} on the striate cortex of the cat and macaque monkey and in the work of Wilder Penfield~\cite{PenfieldandBoldreyBRAIN-37} developing the idea of a cortical homunculus in the primary motor and somatosensory areas of the brain located between the parietal and frontal lobes of the primate cortex. Such maps have become associated with the theory of embodied cognition.}. 
 
As engineers designing the apprentice, part of our job is to create tools that enable the apprentice to learn its trade and eventually become an expert. Conventional IDE tools simplify the job of software engineers in designing software. The fully instrumented IDE (FIDE) that we engineer for the apprentice will be integrated into the apprentice’s cognitive architecture so that tasks like stepping a debugger or setting breakpoints are as easy for the apprentice as balancing parentheses and checking for spelling errors in a text editor is for us.

As a first step in simplifying the use of FIDE for coding, the apprentice is designed to manipulate programs as abstract syntax trees (AST) and easily move back and forth between the AST representation and the original source code in collaborating with the programmer. Both the apprentice and the programmer can modify or make references to text appearing in the FIDE window by pointing to items or highlighting regions of the source code. The text and AST versions of the programs represented in the FIDE are automatically synchronized so that the program under development is forced to adhere to certain syntactic invariants. 


\begin{figure}
  \begin{center} 
    \includegraphics[width=3in]{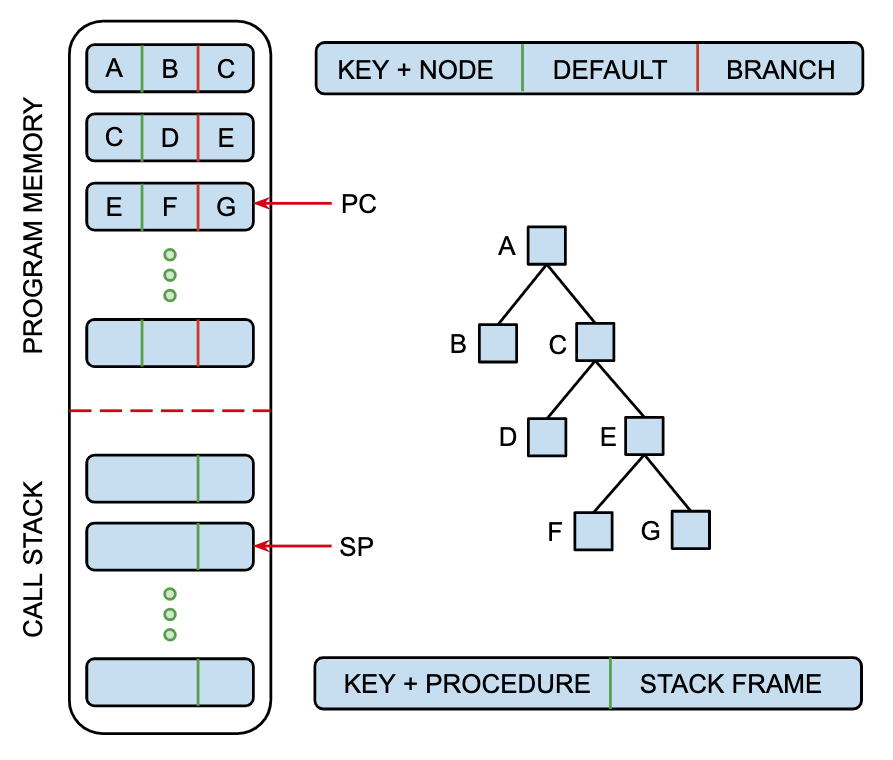} 
  \end{center}
  \caption{We use pointers to represent programs as abstract syntax trees and partition the NTM memory, as in a conventional computer, into program memory and a LIFO execution (call) stack to support recursion and reentrant procedure invocations, including call frames for return addresses, local variable values and related parameters. The NTM controller manages the program counter and LIFO call stack to simulate the execution of programs stored in program memory. Program statements are represented as embedding vectors and the system learns to evaluate these representations in order to generate intermediate results that are also embeddings. It is a simple matter to execute the corresponding code in the FIDE and incorporate any of the results as features in embeddings.}
  \label{fig_programs}
\end{figure}


To support this hypothesis, we are developing distributed representations for programs that enable the apprentice to efficiently search for solutions to programming problems by allowing the apprentice to easily move back and forth between the two paradigms, exploiting both conventional approaches to program synthesis and recent work on machine learning and inference in artificial neural networks. Neural Turing Machines coupled with reinforcement learning are capable of learning simple programs~\cite{GravesetalCoRR-14}. We are interested in representing structured programs expressed in modern programming languages. Our approach is to alter the NTM controller and impose additional structure on the NTM memory designed to support procedural abstraction. 



What could we do with such a representation? It is important to understand why we don’t work with some intermediate representation like bytecodes. By working in the target programming language, we can take advantage of both the abstractions afforded by the language and the expert knowledge of the programmer about how to exploit those abstractions. The apprentice is bootstrapped with several statistical language models: one trained on a natural language corpus and the other on a large code repository. Using these resources and the means of representing and manipulating program embeddings, we intend to train the apprentice to predict the next expression in a partially constructed program by using a variant of imagination-based planning~\cite{PascanuetalCoRR-17}. As another example, we will attempt to leverage NLP methods to generate proposals for substituting one program fragment for another as the basis for code completion. 


\begin{figure}
  \begin{center} 
    \includegraphics[width=2.65in]{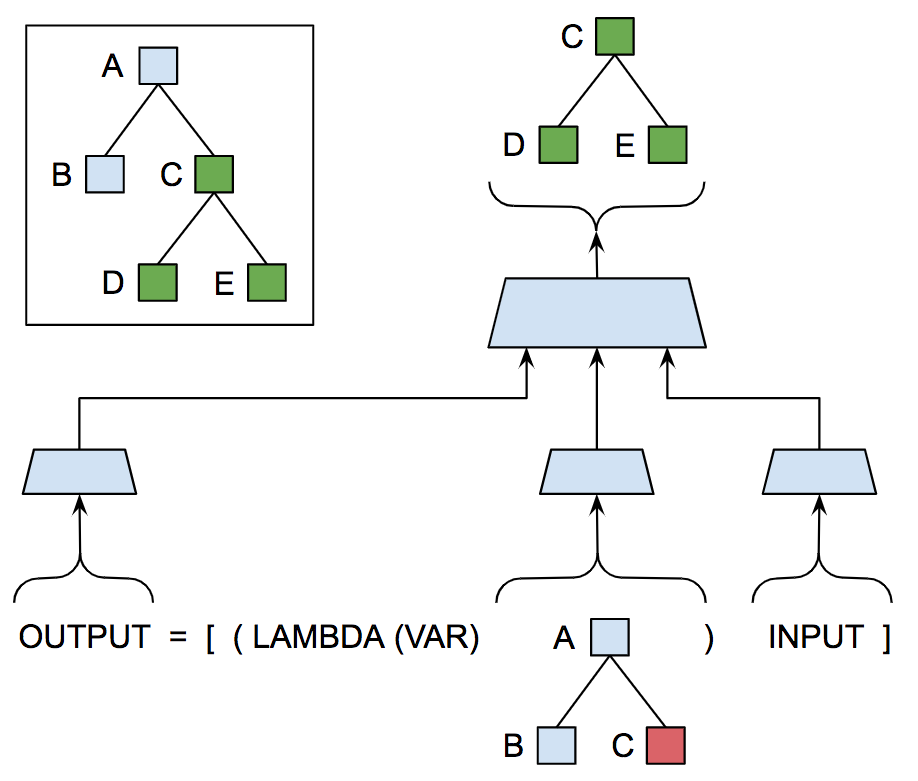} 
  \end{center}
  \caption{This slide illustrates how we make use of input / output pairs as program invariants to narrow search for the next statement in the evolving target program. At any given moment the call stack contains the trace of a single conditioned path through the developing program. A single path is unlikely to provide sufficient information to account for the constraints implicit in all of the sample input / output pairs and so we intend to use a limited lookahead planning system to sample multiple execution traces in order to inform the prediction of the next program statement. 
These so-called imagination-augmented agents implement a novel architecture for reinforcement learning that balances exploration and exploitation using imperfect models to generate trajectories from some initial state using actions sampled from a rollout policy~\cite{PascanuetalCoRR-17,WeberetalCoRR-17,HamricketalCoRR-17,GuezetalCoRR-18}. These trajectories are then combined and fed to an output policy along with the action proposed by a model-free policy to make better decisions. There are related reinforcement learning architectures that perform Monte Carlo Markov chain search to apply and collect the constraints from multiple input / output pairs.}
  \label{fig_emulator}
\end{figure}


The Differentiable Neural Program (DNP) representation and associated NTM controller for managing the call stack and single-stepping through such programs allow us to exploit the advantages of distributed vector representations to predict the next statement in a program under construction. This model makes it easy to take advantage of supplied natural language descriptions and example input / output pairs plus incorporate semantic information in the form of execution traces generated by utilizing the FIDE to evaluate each statement and encoding information about local variables on the stack.



The imagination-based planning (IBP) for reinforcement learning framework~\cite{PascanuetalCoRR-17} serves as an example for how the code synthesis module might be implemented. The IBP architecture combines three separate adaptive components: (a) the {\it{controller}} + {\it{memory}} system which maps a state $s \in S$ and history $h \in H$ to an action $a \in A$; (b) the {\it{manager}} maps a history $h \in H$ to a route $u \in U$ that determines whether the system performs an action in the {\it{compute}} environment, e.g., single-step the program in the FIDE, or performs an imagination step, e.g., generates a proposal for modifying the existing code under construction; the {\it{imagination model}} is a form of dynamical systems model that maps a pair consisting of a state $s \in S$ and an action $a \in A$ to an imagined next state $s' \in S$ and scalar-valued reward $r \in R$.

\begin{figure}
  \begin{center} 
    \includegraphics[width=2.15in]{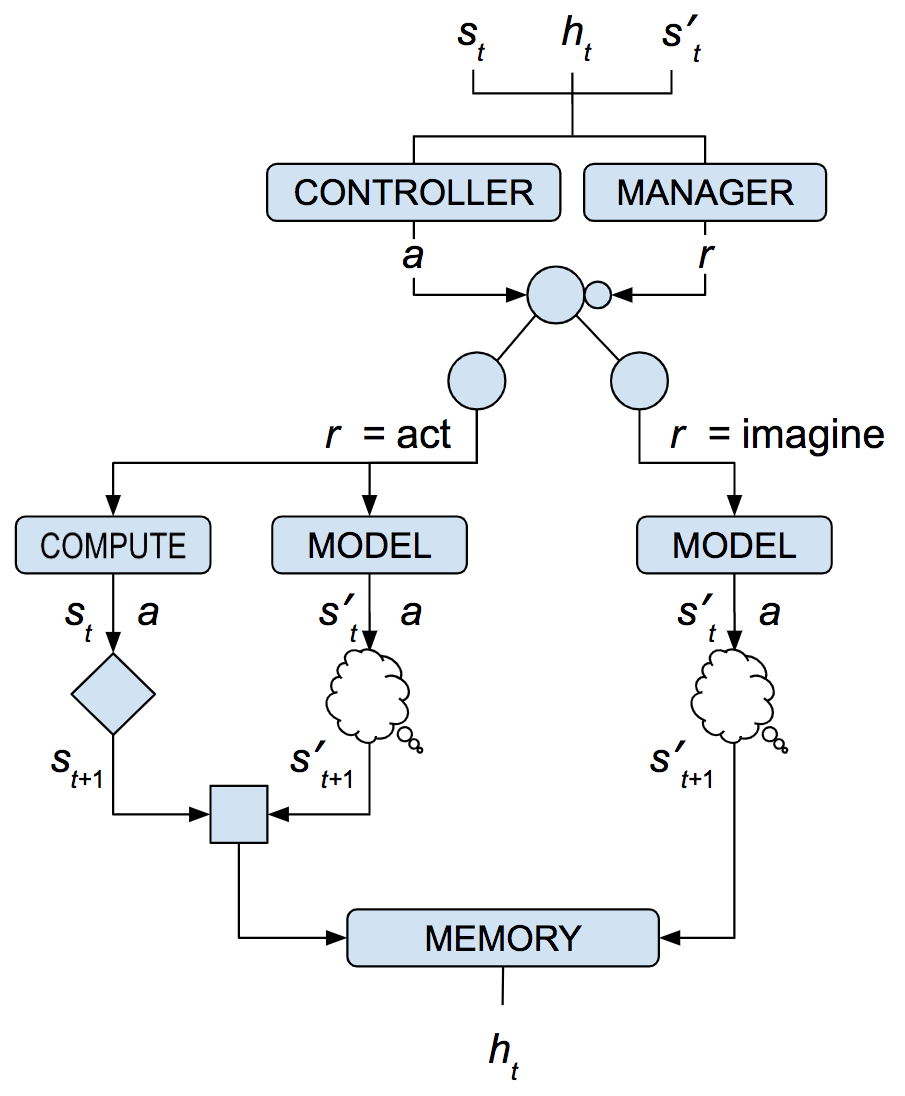} 
  \end{center}
  \caption{The above graphic illustrates how we might adapt the imagination-based planning (IBP) for reinforcement learning framework~\cite{PascanuetalCoRR-17} for use as the core of the apprentice code synthesis module. Actions in this case correspond to transformations of the program under development. States incorporate the history of the evolving partial program. Imagination consists of exploring sequences of program transformations.}
  \label{fig_imagine}
\end{figure}

The imagination model can be implemented as an interaction network~\cite{BattagliaetalNIPS-16} or using the graph-networks framework~\cite{BattagliaetalCoRR-18,SanchezetalCoRR-18}. The three components are trained by three distinct, concurrent, on-policy training loops. The IBP framework shown in Figure~{\urlh{#Graph_Nets_Imagination_Coding}{\ref{fig_imagine}}} allows code synthesis to alternate between exploiting by modifying and running code, and exploring by using the model to investigate and analyze what would happen if you actually did act. The manager chooses whether to execute a command or predict (imagine) its result and can generate any number of trajectories to produce a tree $h_t$ of imagined results. The controller takes this tree plus the compiled history and chooses an action (command) to carry out in the FIDE.


\subsection{End-to-End Systems}
\label{subsection_end-to-end_system}




The subsections that comprise this section of the paper very roughly account for the functional systems of the human brain. Admittedly perception is given short shrift and the systems that deal with emotion and motivation are hardly mentioned at all, the former to save space and the latter because it isn't particularly relevant.

In addition, there was no effort made to map the artificial neural networks described in this section onto the biological subsystems discussed in Section~\ref{section_neuroscience}. The goal was to demonstrate how one might build systems that exhibit some desirable cognitive characteristics of human intelligence by leveraging ideas from neuroscience.

The PBWM ({\it{prefrontal cortex, basal ganglia, working memory}}) model described in~\cite{OReillyandFrankNC-06,HazyetalPTRS-07} covers territory that we only sample from, but the PBWM was developed to explain the function of biological brains, not selectively borrow ideas to extend the capabilities of artificial neural networks.

In a complete end-to-end architecture implementing the programmer's apprentice, the three subsystems described in this subsection would have to be integrated into different parts of the action selection and executive control systems: 
\begin{itemize}
%
\item The system in Figure~{\urlh{#fig_Differentiable_Structured_Programs}{\ref{fig_programs}}} illustrates a differentiable procedural abstraction rich enough to support structured programming in a connectionist setting using standard embedding techniques and memory networks~\cite{WestonetalCoRR-14,DanihelkaetalCoRR-16,GravesetalCoRR-14,GravesetalNATURE-16}.
%
\item The system in Figure~{\urlh{#fig_Differentiable_Program_Emulation}{\ref{fig_emulator}}} provides a sketch of how one might train a language model to predict the next statement in a program under construction using input / output pairs or other program invariants to constrain search~\cite{WangetalCoRR-18,WangetalCoRR-17,SinghandKohliSNAPL-17,DevlinetalICML-17}.
%
\item The system in Figure~{\urlh{#Graph_Nets_Imagination_Coding}{\ref{fig_imagine}}} shows how a variation of imagination-based planning might be used to train a network to predict the next program state using the embodied integrated development environment as a source of ground truth~\cite{WeberetalCoRR-17,PascanuetalCoRR-17,HamricketalCoRR-17}.
\end{itemize}

Exactly how and where these components might be integrated into the overall architecture is beyond the scope of this paper, but research on the neural correlates of mathematical reasoning may provide some useful clues where to start~\cite{DehaeneetalCOGNITIVE-NEUROPSYCHOLOGY-03,DehaeneandBrannon2011mathminds,AmalricandDehaenePNAS-16}.





\section{Discussion}


Training the models described in this paper is a daunting challenge, especially when you consider that current deep neural network technologies rely heavily on large amounts of labeled data and applications like the programmer's apprentice are particularly vulnerable to catastrophic forgetting. We believe training will require radically new approaches and that cognitive and developmental neuroscience have much to offer in terms of insights drawn from the study of how humans learn. 


\subsection{Child Development}


The human brain is organized as a 3-D structure in which specific cell types are positioned in a radial, laminar and areal arrangement that depends on the production, specialization and directed migration of cells from their origin in the embryo to their final destination~\cite{RakiketalTCN-09}. It is only on arriving at their final location that they establish connections to other cells. Postnatally laminar and areal differentiation exhibit substantial differences between early (2-3 months) and late (7-12 months) infancy~\cite{MolnaretalJoA-19,KostovicandJudasTCN-09}. Functional organization begins early (2-3 months) even as construction continues and the shaping of cortical circuits reflects the consequences of increasingly complex behavior. 

All of this carefully orchestrated activity is critical to development. Early brain structures appear as a consequence of the simple reflexive behaviors the infant engages in, laying the foundation for more coordinated behavior depending on increasingly complex internal representations. The infant's ability to engage its environment broadens, exposing it to more complicated stimuli and the opportunity to experiment with new behaviors. The physical and social environment seem to conspire to ensure that the growing infant and then adolescent has the necessary physical and intellectual prerequisites in place when exposed to circumstances that require them. It may be that we will find it useful to recapitulate some version of these developmental strategies for training architectures patterned after the human brain.


\subsection{Inductive Bias}


Most AI systems are trained assuming what is essentially a blank slate in the form of random weights and objective functions that do little to influence the specific content of what is actually learned\footnote{%
  There are exceptions. For example, Ullman~\etal~\cite{UllmanetalPNAS-12} suggest a collection of innate biases that enable the infant visual system to learn to detect human hands by appearance and by context, as well as direction of gaze, in complex natural scenes; Raposo~\etal~\cite{RaposoetalCSC-17} develop models that learn the relational structure of objects and their various arrangements; and Battaglia~\etal~\cite{BattagliaetalCoRR-18} explore the idea of how introducing a relational inductive bias can expedite learning about entities, relations and the rules for their composition.}.
In contrast, babies are born with an innate understanding of how objects move around and interact with one another~\cite{Dehaene-LambertzandSpelkeNEURON-15}. Before they can even crawl about on their own, they appear to have an intuitive understanding of how space, time and number are interrelated~\cite{deHeviaetalPNAS-14,HauserandSpelkeTCN-04}. 
The bodies, neural architectures, physical and social environments of mammals provide a strong inductive bias in shaping their brains. Prolonged development plays a particularly important role in humans. Most mammals can stand and move about within hours of being born. Many human babies don't walk until just under one year. However, human infants learn a great deal during this early preperambulatory developmental period, much of it in preparation for subsequent stages of development~\cite{MacLeanPNAS-16,RosatietalEVOLUTIONARY-PSYCHOLOGY-14}.


It would be difficult if not impossible to genetically encode what a child learns during its lengthy development. And so it seems plausible that evolution would select for a compact general inductive bias enabling us to quickly acquire the basic skills we need to survive while retaining sufficient neural plasticity so that we can adapt to changes during our lifetimes. The schedule of developmental milestones necessary to learn these skills is highly conserved within our species and punctuated by profound changes in the architecture of the brain.

At birth, the architectural foundations are in place to construct the adult brain. For each subsequent developmental milestone, our genes turn on the specific cellular machinery necessary to construct scaffolding, guide neurons of the right cell types to their terminal locations, extend axonal and dendritic processes, eliminate unnecessary neurons and establish new or prune existing synaptic connections. The innate inductive bias and training curriculum implicit in development influence two critical factors that determine human intelligence: First, they serve to initialize the mapping from body to latent state representations throughout the cortex thereby grounding experience in the physical environment. Second, they utilize this grounding as the basis for all subsequent understanding, concrete and abstract. 

This basis provides a template or prototype\footnote{%
  Prototype theory is a mode of graded categorization in cognitive science, where some members of a category are more central than others. For example, when asked to give an example of the concept furniture, chair is more frequently cited than, say, stool. Prototype theory has also been applied in linguistics, as part of the mapping from phonological structure to semantics. ({\urlh{https://en.wikipedia.org/wiki/Prototype_theory}{SOURCE}})}
for representing new concepts, whether they be predictive models that allow us to interact with complex dynamical systems or composite categorical representations that enable us to recognize, contrast and compare instances of a particular class of entities~\cite{RoschNATURAL-CATEGORIES-95,RoschNATURAL-CATEGORIES-91,VarelaThompsonRoschTHE_EMBODIED_MIND-91}. 
In designing architectures to accommodate learning such representations, recent work on learning relational models that characterize different classes of entities, the relationships they participate in and the rules employed in composing them to form new relationships seems particularly promising~\cite{SanchezetalCoRR-18,HamricketalCoRR-18,SantoroetalNIPS-17,BattagliaetalNIPS-16}. This core competency should also serve as the starting point for reasoning about all sorts of abstract entities including computer programs and mathematical objects\footnote{%
  Learning entities, relations and their compositions may seem like it would require exotic new architectures, but the work of Hubel and Wiesel~\cite{HubelandWieselJoP-62,HubelandWieselJoP-61} published nearly sixty years ago that inspired Fukushima~\cite{FukushimaBC-80}, LeCun~\etal{}~\cite{LecunetalIEEE-98} and Riesenhuber and Poggio~\cite{RiesenhuberandPoggioNN-99} among many others is relevant to much more than extracting features from images.\\
  Convolutional neural networks are common in many applications areas including natural language processing~\cite{KimEMNLP-14}, medical data analysis~\cite{Luetal2017medical}, extreme weather condition prediction~\cite{LiuetalCoRR-16}, time series prediction for speech modeling~\cite{vandenOordCoRR-16} to name but a few. Many of these applications combine CNN and RNN technologies using CNN components to model coarse-grained local features generated and RNN models to account for long-distance dependencies~\cite{WangetalCOLING-16}.\\
  Convolutional layers in a multilayer stack employing a sparsity-inducing energy function essentially segment their input providing a basis for entity recognition. Subsequent pooling layers serve to identify correlations evident in earlier layers and thereby identify candidates for relationships, potentially including pairwise and higher-order relationships in a deep enough stack~\cite{SantoroetalNIPS-17}. In addition to Hubel and Wiesel, this technology owes a debt to Barlow~\cite{BarlowNC-89,Barlow61}, Rao and Ballard~\cite{RaoandBallardNATURE-NEUROSCIENCE-99} for foundational biologically-inspired work on sparse and predictive coding~\cite{ChalketalPNAS-18}.}.


\subsection{Natural Language}


The use of language and, more generally, symbolic reasoning is an important if not defining characteristic of human intelligence. Some cognitive scientists, including such outspoken proponents as Jerry Fodor and Zenon Pylyshyn, view symbolic representations that exhibit combinatorial syntactic and semantic structure as candidates for a language of thought, and view connectionist proposals as lacking these properties and serving primarily as an account of the neural structures in which symbolic representations are implemented~\cite{FodorandPylyshynCOGNITION-88,Fodor84}. O'Reilly~\etal{}~\cite{OReillyetalTACO-14} have attempted to reconcile the symbolic and connectionist views, arguing that the two are complementary and that parts of the brain exhibit properties of both symbolic and connectionist information processing. 

The evolutionary biologist, Terrence Deacon, argues that our use of language is not a direct consequence of natural selection but rather the result of a collective effort involving millions of human beings working over thousands of year to produce an encyclopedic record of human endeavor to pass down to future generations~\cite{Deacon1998symbolic}. It is hard to imagine an effective programmer's apprentice, much less an accomplished software engineer, lacking the ability to communicate in natural language or denied access to the written word. Recent progress in grounding language learning in an agent's experience interacting with a suitably complex environment bodes well for applications like the programmer's apprentice. For these reasons and more, we see the rich complexity of collaborative pair programming as a compelling framework for exploring human-level AI.

\end{document}